\documentclass[]{aastex62}
\usepackage{amssymb,amsmath}

\def\eq#1{\begin{equation} #1 \end{equation}}
\def\mic              {\hbox{$\mu\mathrm{m}$}}
\def\atm    {\hbox{\rm{ATM}}}

\graphicspath{{./}{compressed/}}

\received{May 7, 2019}
\revised{May 7, 2019}
\accepted{--}
\submitjournal{Icarus}

\shorttitle{\atm: An Open-Source Tool for Asteroid Thermal Modeling}
\shortauthors{Moeyens, Myhrvold \& Ivezi\'{c}}

\begin{document}

\title{{\rm ATM}: An Open-Source Tool for Asteroid Thermal Modeling}

\correspondingauthor{Joachim Moeyens}
\email{moeyensj@uw.edu}

\author[0000-0001-5820-3925]{Joachim Moeyens}
\affiliation{Department of Astronomy and the DIRAC Institute, University of Washington, 3910 15th Avenue NE, Seattle, WA 98195, USA}
\affiliation{LSSTC Data Science Fellow}

\author{Nathan Myhrvold} 
\affiliation{Intellectual Ventures, Bellevue, WA 98005, USA} 

\author[0000-0001-5250-2633]{\v{Z}eljko Ivezi\'{c}}
\affiliation{Department of Astronomy and the DIRAC Institute, University of Washington, 3910 15th Avenue NE, Seattle, WA 98195, USA}

\begin{abstract}
We publicly release \atm, a Python package designed to model asteroid flux measurements to estimate an asteroid's size, surface temperature distribution, and emissivity. 
A number of the most popular static asteroid thermal models are implemented with the reflected solar light contribution and Kirchhoff's law accounted for. Priors for fitted parameters 
can be easily specified and the solution, including the full multi-dimensional posterior probability
density function, is found using Markov Chain Monte Carlo (MCMC). We describe the package's
architecture and discuss model and fitting validation. Data files with $\sim$ 2.5 million WISE flux measurements 
for $\sim$ 150,000 unique asteroids and additional Minor Planet Center data are also included with the 
package, as well as Python Jupyter Notebooks with examples of how to select subsamples of 
objects, filter and process data, and use \atm\ to fit for the desired model parameters. The entirety of the analysis
presented here, including all the figures, tables, and catalogs, can be easily reproduced with these publicly released Notebooks. We show that \atm\ can match the best-fit size estimates for well-observed asteroids published in 
2016 by the NEOWISE team \citep{2016PDSS..247.....M}
with a sub-percent bias and a scatter of only 6\%. Our analysis of various sources of random and 
systematic size uncertainties shows that for the majority of over 100,000 objects with WISE-based 
size estimates random uncertainties (precision) are about 10\%;  systematic uncertainties within the adopted 
model framework, such as NEATM, are likely in the range of 10-20\%. Hence, the accuracy of WISE-based 
asteroid size estimates is approximately in the range of 15-20\% for most objects, except for unknown errors due 
to a possibly over-simplified modeling framework (e.g., spherical asteroid approximation).  We also 
study optical data collected by the Sloan Digital Sky Survey (SDSS) and show that correlations of 
optical colors and WISE-based best-fit model parameters indicate robustness of the latter. Our analysis  
gives support to the claim by \cite{2014ApJ...785L...4H} that candidate metallic asteroids can be selected 
using the best-fit temperature parameter and infrared albedo. We investigate a correlation between SDSS colors  
and optical albedo derived using WISE-based size estimates and show that this correlation can 
be used to estimate asteroid sizes with optical data alone, with a precision of about 21\% 
relative to WISE-based size estimates. After accounting for systematic errors, the difference in accuracy
between infrared and optical color-based size estimates becomes less than a factor of two. 
\end{abstract}


\section{Introduction}

Asteroid thermal flux modeling aims to estimate an asteroid's size (volume-equivalent diameter,
hereafter diameter) and surface temperature distribution,
and sometimes other physical properties such as emissivity, from measured infrared (IR) fluxes. The largest 
dataset of infrared flux measurements for asteroids was recently contributed by the WISE survey \citep{WISE}
and analyzed by the associated NEOWISE team \citep[][and references therein]{NEOWISE}. Flux measurements in four WISE bands
provide strong constraints on asteroid sizes and emissivities 
as a function of wavelength. A series of papers that produced size estimates for about 164,000 asteroids, 
as well as constraints on asteroid emissivity properties, was reviewed and summarized by \cite{MainzerReview}. 

It appears that the pioneering analysis by the NEOWISE team can be improved in various ways, 
as argued by \cite{NM2018a, NM2018b}. In particular, ignoring reflected sunlight can induce 
biases in estimated asteroid sizes and lead to underestimated size uncertainties. In addition, 
best-fit sizes can be biased due to assumptions on the priors for an asteroid's emissivity as a function of wavelength, which varies with the chemical composition of an asteroid's surface. Further biases are introduced by the use of fairly simplistic static thermal models to calculate volume-equivalent diameters. 
The need for improvements in data analysis was recently acknowledged by \cite{2018arXiv181101454W}, but it appears that a number of data analysis issues remain open. In particular, the behavior of systematic and random (statistical) uncertainties for the best-fit parameters remains an active research topic 
\cite[see for example][]{2007astro.ph..3085W, 2018AJ....155...74M, 2018AJ.156.62M}

Given that the WISE dataset is by far the largest of its kind and will not soon be surpassed, it is prudent to focus on improved data analysis. We aim to contribute to
such improvements by publicly releasing a new Python modeling tool, \atm\ (Asteroid
Thermal Modeling\footnote{See \url{https://github.com/moeyensj/atm}}), designed to enable easy fitting of the most common thermal models to WISE and other asteroid flux measurements. Data analysis software is often a crucial component in delivering scientific results, as vividly exemplified by this case,
and thus discussions of scientific reproducibility and transparency can be greatly enhanced by collaborative software development and code sharing. Thanks to rapidly developing tools and technologies,
such as Jupyter Notebooks, Python, and GitHub, performing these steps in 
open-source environment is now easier than ever. By releasing \atm\ we
aim to increase reproducibility -- a fundamental tenet of 
the scientific process. 
 
In \S2 we describe the mathematical/physical model underlying \atm, its Python implementation,
and discuss model validation using observational and model data for asteroid sizes from the literature. 
The capabilities of \atm, with emphasis on various treatments of Bayesian priors when also fitting 
emissivity are further illustrated in \S3 using three well-observed asteroids. 
In \S4 we apply \atm\ to a ``gold'' sample of $\sim$7,000 best-observed asteroids from the NEOWISE 
dataset; in addition to best-fit diameters, we also obtain best-fit values for the emissivity of each asteroid
across WISE bands W1 and W2. The best-fit sizes are compared to the values obtained by the 
NEOWISE team (the 2016 Planetary Data System version\footnote{See \url{https://sbn.psi.edu/pds/resource/neowisediam.html}}).
In \S5 we match the ``gold'' sample to optical data from the SDSS Moving Object Catalog, compute the visual albedo $p_V$, and study
its correlation with optical colors. We also discuss how optical colors can be used to 
estimate $p_V$ and asteroid sizes when adequate infrared data are not available. Our results
are summarized in \S6. 

As part of the \atm\ package release, we also include all data files used in this work including NEOWISE, 
Minor Planet Center, and SDSS Moving Object Catalog data. We provide Jupyter Notebooks with examples 
of how to select subsamples of objects, process and filter their flux measurements, 
and how to use \atm\ to estimate their diameters and infrared emissivities. All of the analysis 
presented here, including all the figures and tables, can be easily reproduced using the Notebooks 
released with the \atm\ package on GitHub. In particular, \url{https://github.com/moeyensj/atm/blob/master/notebooks/README.md} lists the Notebooks required to reproduce each figure in this paper.

\section{ATM: An Open-Source Tool for Asteroid Thermal Modeling}

A detailed discussion of the relevant physics implemented in asteroid thermal flux modeling,
and a summary of models proposed in the literature, are presented in \cite{NM2018a} and references therein. Here we only summarize the main results needed to understand how \atm\ works and what it computes. We also describe its Python implementation and discuss model 
validation using observational and model data for asteroid sizes from literature. 

\subsection{The Asteroid Flux Model Summary} 

We first introduce $F_\nu(\lambda)$: the specific flux (flux per unit frequency, $\nu$) 
from an object. The SI units for $F_\nu$ are W m$^{-2}$ Hz$^{-1}$ (= 10$^{3}$ erg cm$^{-2}$ s$^{-1}$). 
The specific flux can also be defined per unit wavelength, $F_\lambda$, using energy
conservation $F_\nu |d\nu|=F_\lambda |d\lambda|$ and $\lambda \nu = c$.
The choice of $F_\nu$, as opposed to $F_\lambda$, is completely arbitrary. Similarly,
the running variable can be either $\lambda$ or $\nu$, and the choice of $\lambda$ is 
more convenient in this context. 

The model flux from an asteroid, $F_\nu^{ast}(\lambda)$, corresponding to flux detected by the observer, $F_\nu^{obs}(\lambda)$, is the 
sum of the emitted thermal flux controlled by the asteroid's surface temperature distribution,
and the portion of the incident solar flux reflected by the asteroid, 
\eq{
     F_\nu^{ast}(\lambda) = F_\nu^{th}(\lambda) + F_\nu^{ref}(\lambda). 
}
A given model spectrum $F_\nu^{ast}(\lambda)$, corresponding to observational quantity $F_\nu^{obs}(\lambda)$,
is integrated over the bandpass (assumed known hereafter) to obtain observed in-band model flux for a given 
instrument. For example, \cite{2013AAS...22143905W} has derived simple quadrature formulae\footnote{See Appendix A 
for a slight correction to the quadrature formula for the W3 band.} that can be used to 
efficiently and accurately compute in-band fluxes for the four WISE bands from model flux 
$F_\nu^{ast}(\lambda)$.

Both $F_\nu^{th}(\lambda)$ and $F_\nu^{ref}(\lambda)$ depend on the relative positions
of the Sun, the asteroid and the observer, the asteroid's diameter $D$, and the asteroid's emissivity, $\epsilon(\lambda)$, which controls the balance between absorbed/emitted and reflected incident flux. These relative positions
are fully described by the asteroid-Sun distance, $r$, asteroid-observer distance,
$\Delta$, and the angle subtended on the asteroid's surface by the lines of sight towards 
the Sun and the observer, the so-called phase angle, $\alpha$. Hereafter we assume 
that asteroid orbital parameters are known and that $r$, $\Delta$ and $\alpha$ can
be easily computed using standard and readily-available tools \cite[e.g., the JPL HORIZONS service\footnote{See \url{https://ssd.jpl.nasa.gov/?horizons}}; OpenOrb, ][]{2015ascl.soft02002G}. To simplify nomenclature, we do not explicitly list these independent variables, unless 
necessary to avoid confusion.

\subsubsection{The emitted flux \label{sec:emittedFlux}} 

The observed thermal flux is obtained by integrating the emitted thermal flux
per unit area over the visible surface of the asteroid, 
\eq{
\label{eq:thermalF}
F_\nu^{th}(\lambda) = \left( { D \over 2\, \Delta}\right)^2 \, 
\epsilon(\lambda) \int_{-\pi/2}^{\pi/2} \int_{-\pi/2+\alpha}^{\pi/2+\alpha} \pi 
B_{\nu}\left(T(\theta, \phi), \lambda \right) \cos(\theta-\alpha) \cos^2(\phi) d\theta  d\phi,
}
where $\theta$ and $\phi$ are the integration 
variables over the asteroid's surface (here we use geographic coordinates with $\theta=0$ and $\phi=0$ at the subsolar point,
and both ranging from $-\pi$ to $\pi$; note that \citealt{NM2018a} used ISO coordinates with $0 < \phi <2\pi$),
and $B_{\nu}$ is the Planck function. 

It is implied that, in this context of thermal emission, asteroids are approximated as perfect spherical Lambertian emitters which follow Lambert's cosine rule (unlike scattered light, which shows a strong opposition surge -- a peaked reflectance for phase angles 
near zero). 

The temperature variation across an asteroid's surface is model dependent.  \atm\ implements 
the three most common thermal models: the Standard Thermal Model (STM), the Fast Rotating Model 
(FRM), and the Near-Earth Asteroid Thermal Model (NEATM). For a comparison of these models and an analysis of their validity, please see \cite{2007astro.ph..3085W}, \cite{2018AJ....155...74M}, and references therein. Common to
all models is a temperature scale set by the so-called sub-solar temperature, $T_{ss}$, 
which is the highest temperature on the asteroid's surface. Depending on the adopted 
model, the temperature variation across the surface can range in complexity from being 
constant to having a strong temperature gradient from the subsolar point (e.g., see Fig. 1 in \citealt{2018AJ....155...74M}).

The NEATM model uses a multiplicative correction factor (e.g.,
for surface roughness) to eq.~\ref{eq:thermalF} called the``beaming parameter'', $\eta$. For a more detailed discussion of this parameter, please see \cite{NM2018a}. 
The $\eta$ parameter is not constant and can vary with the observer's
location \citep{2007astro.ph..3085W}.

Given a model parameter for the temperature scale, $T_1$, defined as the sub-solar 
temperature when $r=1$ A.U., the sub-solar temperature is simply
\eq{
     T_{ss} = \left( { 1 \, {\rm A.U.} \over r}\right)^{1/2} \, T_1. 
}
Unlike $T_{ss}$, which depends on the asteroid-Sun distance, $T_1$ is nearly constant for a given asteroid
(not exactly constant because of its slight dependence on $\eta$, see eq.~\ref{eq:T1} below).

The energy balance, discussed below, connects $T_1$ with the incident solar flux, an
asteroid's physical properties and other model parameters. \cite{NM2018a} made an 
important point that when fitting a model to data, all the other parameters are
not directly relevant -- it is only $T_1$ that controls the model fluxes and thus it is 
only $T_1$ that is directly constrained by the infrared flux data for a given observer's position. 

\subsubsection{The energy balance and the meaning of best-fit parameter $T_1$} 

Using the energy balance equation that equates the absorbed incident solar flux 
with the flux emitted by an asteroid, it can be shown (e.g., see \citealt{NM2018a} 
and references therein) that 
\eq{
\label{eq:T1}
T_1 = \left( {S (1-A) \over \sigma \eta \epsilon_B} \right)^{1/4}, 
}
where $S$ is the solar constant at 1 A.U. ($S = 1360.8$ Wm$^{-2}$), 
$\sigma=5.67\times10^{-8}$ W m$^{-2}$ K$^{-4}$ is the Stefan-Boltzmann constant, 
and $A$ and $\epsilon_B$ are appropriately wavelength-averaged values of $1-\epsilon(\lambda)$ and $\epsilon(\lambda)$ over the incident and thermal flux distributions, respectively (for details, 
please see \S3 in \citealt{NM2018a}). It is usually assumed that $\epsilon_B=0.9$,
but the exact value is not crucial to modeling once the beaming parameter $\eta$
is introduced -- it is only the $\eta \epsilon_B$ product that can be constrained
using best-fit $T_1$. Another conventional approximation is 
\eq{
        A \approx  A_V = p_V \, q,
}
where $A_V$ is the Bond albedo (limited to the range 0--1), $p_V$ is the geometric albedo in the visible band (can be larger than 1), and $q$ is the empirically-derived phase integral. In the H-G magnitude system 
introduced by \cite{Bowell1989}, $q(G) = 0.29 + 0.684G$, where $G$ is the slope 
parameter of the phase function.  
A common assumption, when G is unknown, is $G=0.15$. Therefore, the best-fit $T_1$ can be used to 
estimate $p_V$, given the value of the $\eta \epsilon_B$ product.  

The geometric albedo can also be constrained, when the asteroid's diameter, $D$, is known,
using flux measurements at wavelengths sufficiently short for flux to be dominated by the reflected incident solar flux. For example, in the visual band
\eq{
\label{eq:pV}
         p_V = \left( { 1,329 \, {\rm km} \over D} \right)^2 \, 10^{-0.4 H},
}
where $H$ is the asteroid's absolute magnitude in the visual band. When such 
measurements are available, the two constraints for $p_V$ can be used to estimate the 
beaming parameter, $\eta$, or to simply check the model's internal consistency.

\subsubsection{The reflected flux} 

The reflected flux is proportional to an asteroid's reflectivity, $\rho(\lambda)$, with
$\rho(\lambda) = 1 - \epsilon(\lambda)$ via Kirchhoff's law, 
\eq{\label{eq:Fref}
    F_\nu^{ref}(\lambda) =  \left( { D \over 2\, \Delta}\right)^2 \, 
    \frac{\Psi(\alpha, G)}{q(G)} \,\left[1 - \epsilon(\lambda)\right] \, F_\nu^\sun(\lambda),
}
where the H-G phase function, $\Psi(\alpha, G)$, and empirically derived phase integral, $q(G)$, are purely geometric quantities that account for phase effects \citep{Bowell1989}.

The incident solar flux, $F_\nu^\sun(\lambda)$, at a distance $r$ from the Sun, is given by
\eq{
    F_\nu^\sun(\lambda) = \left({R_\sun \over r}\right)^2 \, \pi \, B_\nu(T_\sun, \lambda),
}
where $T_\sun = 5,778$ K and $R_\sun=0.00465$ A.U. Therefore, when $\epsilon(\lambda)$ is known, the reflected flux is fully determined for a given observing geometry (note also that 
a fully known $\epsilon(\lambda)$ implies a given value of $p_V$). 

\subsubsection{Fitting Model Fluxes to Data} 

The likelihood of obtaining $N$ observed fluxes, $F^{obs}_i$, given model predictions, $F_i^{ast}$, 
can be written as (for details see, e.g., Chapter 4 in \citealt{zeljkoBook})
\eq{
\label{eq:MLflux}
    L =\prod_{i=1}^N  {1 \over \sqrt{2\pi (\sigma_i^2+\Sigma^2)}} \exp{\left({-(F^{obs}_i-F_i^{ast})^2 \over 
        2 \, (\sigma_i^2 + \Sigma^2)}\right)},
}
where $\sigma_i$ is the flux measurement uncertainty and $\Sigma$ accounts for variability. 
Assuming flat Bayesian priors for fitted model parameters, maximizing this likelihood 
function is equivalent to maximizing the Bayesian posterior probability density function. \atm\ also supports the Jeffreys’ 
priors (flat distributions in $\log(D)$ and $\log(T_1)$, for details 
see Chapter 5 in \citealt{zeljkoBook}). 

The above expression assumes that the scatter of flux measurements around predicted 
model values follows a Gaussian distribution with standard deviation (measurement 
uncertainty) $\sigma_i$. In practice, most asteroids show variability with amplitudes
($\sim 1$ mag) exceeding typical measurement uncertainties ($\le$0.2 mag for SNR$\ge 5$). This variability
is mostly due to non-spherical shapes, which are not captured by the model. For this
reason, the likelihood expression includes the $\Sigma^2$ term. Given that 
$\Sigma \gg \sigma_i$ for reasonably high SNR, we have $\sigma_i^2 + \Sigma^2 \approx \Sigma^2$, which
is presumed constant for a given asteroid. While it does not influence the values of 
the best-fit parameters, $\Sigma$ does control their uncertainty. Following \cite{2018AJ....156...60M}, 
we adopt $\Sigma=0.2$ mag as typical uncertainty due to variability. 

When emissivity $\epsilon(\lambda)$ is assumed known, the best-fit model parameters are effectively 
obtained by maximizing the log-likelihood in magnitude space, as a function of two free 
parameters, $D$ and $T_1$ (for a discussion of emissivity fitting, see the next section),
\eq{
\label{eq:MLmag}
  ln(L) = {\rm const.} - \sum_{i=1}^N \left[m^{obs}_i-m_i^{ast}(D,T_1) \right]^2,
}
where $m^{obs}_i$ are observed magnitudes and $m_i^{ast}$ are their model predictions ($m = -2.5\,\log_{10}(F) + {\rm const.}$). 

\subsubsection{The treatment of emissivity, $\epsilon(\lambda)$ \label{sec:eps}} 

The assumed priors for $\epsilon(\lambda)$ play an important
role in determining an asteroid's size. Technically, all fitted parameters require their prior probability distributions to be specified (for a discussion of Bayesian priors, see, for example, Chapter 5 in \citealt{zeljkoBook}). When $\epsilon(\lambda)$ is assumed known, these priors effectively become the Dirac $\delta$ functions. 

When $\epsilon(\lambda)$ is assumed known, in principle even just two infrared flux measurements can constrain an asteroid's diameter $D$ and temperature $T_1$. In essence, the measured
color (i.e. the ratio of the two measured fluxes) constrains $T_1$ and the overall flux level constrains $D$. When measurement errors are present, the constraints become more complex, as discussed further below in the model validation section (\S\ref{sec:valid}). If more than two infrared flux measurements are available, $\Sigma$ is known from light curve analysis, and measurement errors are sufficiently small, the model fluxes can be checked for internal consistency using statistical tests such as $\chi^2$. 

When $\epsilon(\lambda)$ is treated as a free parameter,
care must be taken to avoid the $D\epsilon$ degeneracy when
only thermal fluxes are measured: in eq.~\ref{eq:thermalF}
parameters $D$ and $\epsilon$ appear only as a product $D^2\epsilon$.
This degeneracy can be broken when flux measurements at wavelengths where the reflected light component is non-negligible
are available. Because in eq.~\ref{eq:Fref} the reflected flux is proportional to $D^2(1-\epsilon)$ rather than proportional to
$D^2\,\epsilon$, the $D-\epsilon$ degeneracy is broken
and both $\epsilon$ and $D$ can be estimated.
The larger the contribution of the reflected flux to the
total flux, the less correlated are the best-fit values
of $\epsilon$ and $D$. We return to this point in \S\ref{sec:valid}. 

In reality, a rather substantial variation in $\epsilon(\lambda)$ is observed for asteroids (e.g., 
see Fig. 1 in \citealt{NM2018a}). If more than two flux measurements are available, as is
the case for NEOWISE data with four bands, parameters from a judiciously chosen 
parameterization of $\epsilon(\lambda)$ can also be fit to data. In the case of four bandpasses,
up to two such parameters can be well constrained\footnote{Here we ignore statistical techniques 
such as regularization, as well as hierarchical Bayesian modeling, which can be used to fit 
for more parameters than there are data points; for more details, please see Chapter 8 in
\citealt{zeljkoBook}.} (in addition to $D$ and $T_1$, for the total 
of four fitted parameters). 

Unknown functions are often specified as power laws with two free parameters: the 
power-law index and its overall normalization. This common approximation would work poorly 
given the $\epsilon(\lambda)$ functions observed for asteroids. Instead, motivated by the
observed behavior of the $\epsilon(\lambda)$ functions, we consider an ansatz (``a simplified
procedure'', or ``a provisional mathematical assumption'') where 
we fit $\epsilon_{W1W2}$, the value of $\epsilon(\lambda)$ in the WISE bands 1 and 2, and
$\epsilon_{W3W4}$, the value of $\epsilon(\lambda)$ in the WISE bands 3 and 4. In other words,
we approximate $\epsilon(\lambda)$ as a step function $\epsilon(\lambda) = \epsilon_{W1W2}$ 
for $\lambda < 6 \mic$, and $\epsilon(\lambda) = \epsilon_{W3W4}$ for $\lambda > 6 \mic$. Because the two shortest WISE bands include substantial contribution of reflected flux (e.g., see
Figure 1 in \citealt{NM2018a}), constraints can be placed on $\epsilon_{W1W2}$ 
and $\epsilon_{W3W4}$ for well-observed asteroids.
We discuss these constraints in more detail in the next section.
We note that asteroids with sizes estimated by other means
(e.g., from occultation or radar measurements) represent
an invaluable sample for constraining the behavior of 
$\epsilon(\lambda)$ among the asteroid population.

\subsection{Model Implementation: \atm} 

We now briefly describe the \atm\ code architecture and highlight some aspects of its functionality. \atm\ is built on two bespoke Python classes: the `observatory' and `model' classes. 
The observatory class is used to describe an observatory's filter throughput curves as a function of wavelength. 
In the case of WISE, these are the modified quadrature formulae provided by \cite{2013AAS...22143905W}.
The model class has placeholder functions for how a thermal model describes the surface temperature distribution on a model asteroid, how the observing geometry is taken into account, and how the model integrates emitted flux over the surface of a model asteroid of unit diameter. 
The instances of the two classes, called objects, can then be passed as arguments with physical parameters ($D$, $T_{ss}$, $\epsilon$, $p$, $G$) and observing parameters ($\alpha$, $r$, $\Delta$) to a series of functions. These functions calculate predicted in-band fluxes with and/or without reflected sunlight for the given observatory. 
Additionally, the model class can be passed separately to a different series of functions to generate model 
spectral energy distributions (SEDs) at a range of desired wavelengths. 

As indicated by \cite{NM2018a}, fitting for different desired parameters and integrating the emitted flux on the fly can be slow. 
To make modeling and subsequent fitting more computationally tractable, \atm\ comes with emitted flux lookup tables generated for each thermal model at each WISE quadrature wavelength and at a range of wavelengths at 1 micron intervals between 1 and 30 microns.
The lookup tables are functions of a grid of $T_{ss}$ and $\alpha$ values and contain $\sim$ 800,000 saved integration evaluations per table.

To conduct MCMC sampling, \atm\ uses the \textit{pymc3} \citep{pymc3} package. \textit{pymc3} is a robust Python probabilistic programming and Bayesian inference package and it allows for multi-threaded sampling of the posterior; the \atm\ lookup table implementation in combination with \textit{pymc3} allows 4,000 samples to be extracted in approximately 40 seconds on a moderate CPU. Each sampling chain requires a single thread, hence, sampling the posterior with 10 chains would require 10 threads and, because \textit{pymc3} support parallelization, it takes the same amount of time to sample as a single chain.
Given a set of asteroid flux measurements for either a single object or multiple objects, with assumptions on priors for fitting parameters, the user can fit for any combination of parameters as defined by Eq \ref{eq:thermalF}. 
Should the user decide to constrain emissivity as described in \S\ref{sec:eps}, this task too is made simple by the use of several keyword arguments in the fitting function that describe how emissivity and albedo should be constrained and calculated. 

We refer the reader to the GitHub repository for further information on the modeling code. 
The Notebooks used to validate the code are located here: \url{https://github.com/moeyensj/atm/tree/master/notebooks/validation/}. 
The Notebooks used to analyze the WISE sample of data (presented in the following sections) are located here: \url{https://github.com/moeyensj/atm/tree/master/notebooks/analysis/}. 
A Notebook showing how different assumptions regarding emissivity and albedo affect the modeling and retrieval of physical parameters is located here: \url{https://github.com/moeyensj/atm/tree/master/notebooks/analysis/single_object_90367.ipynb}.  

\subsection{\atm\ Validation \label{sec:valid}} 

No complex software system can be fully trusted without extensive validation and verification steps. 
We validate and verify the \atm\ code in two steps. First, we generate synthetic fluxes from a 
hypothetical asteroid using given $D$, $T_1$, and $\epsilon(\lambda)$, and 
then use \atm\ to perform Markov Chain Monte Carlo (MCMC) fitting to recover 
the input parameters. This step serves as a validation of internal
code consistency. Second, physical validation is performed using
a few studies from the literature, where asteroid sizes are
known independently from infrared flux measurements (e.g., from radar 
measurements and stellar occultation measurements). 

\subsubsection{Validation of the internal code consistency}\label{sec:internalvalid}

\begin{figure}[t]
\centering
\includegraphics[width=0.9\textwidth, keepaspectratio]{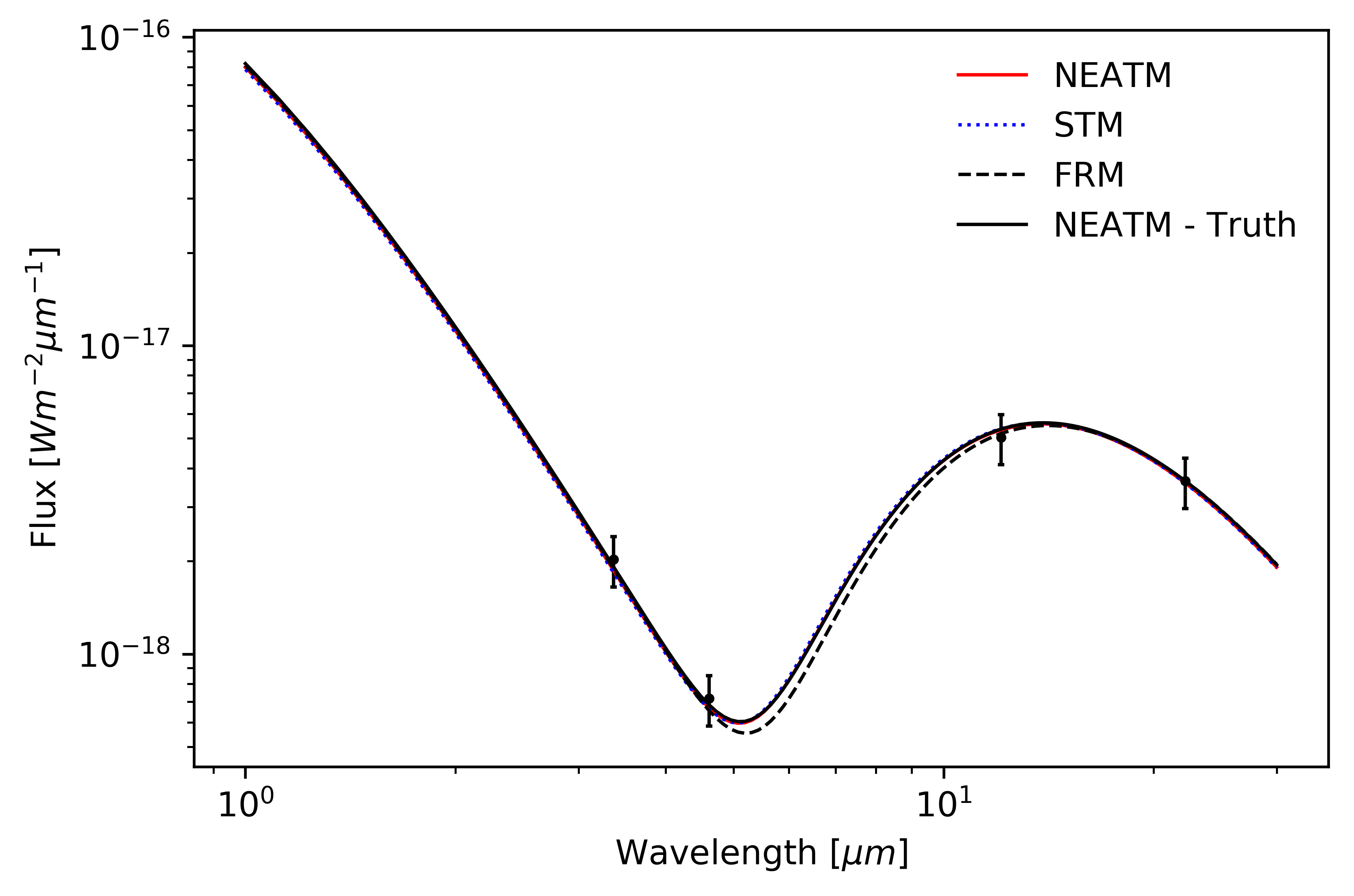}
\caption{The solid line shows the synthetic spectral energy distribution (SED)
generated using the NEATM model for an asteroid with $D = 1$ km, $T_1 = 422$ K, 
$\epsilon(\lambda)=0.7$, asteroid-Sun distance of 3 A.U., and an observer
at 2 A.U. The four symbols with error bars correspond to the four in-band fluxes 
for the WISE observatory at the same distance from the asteroid. The error bars reflect the assumed variability 
amplitude but the implied scatter is not shown in this plot. The three other
lines show the best fits obtained with \atm\ (see text). The two solid lines are 
indistinguishable. This figure was generated using \href{https://nbviewer.jupyter.org/github/moeyensj/atm/blob/master/notebooks/validation/example_synthetic_changingNumObservations.ipynb}{notebooks/validation/example\_synthetic\_changingNumObservations.ipynb}.
\label{fig:synthFlux}}
\end{figure}
 
\begin{figure}[t]
\centering
\includegraphics[width=0.9\textwidth, keepaspectratio]{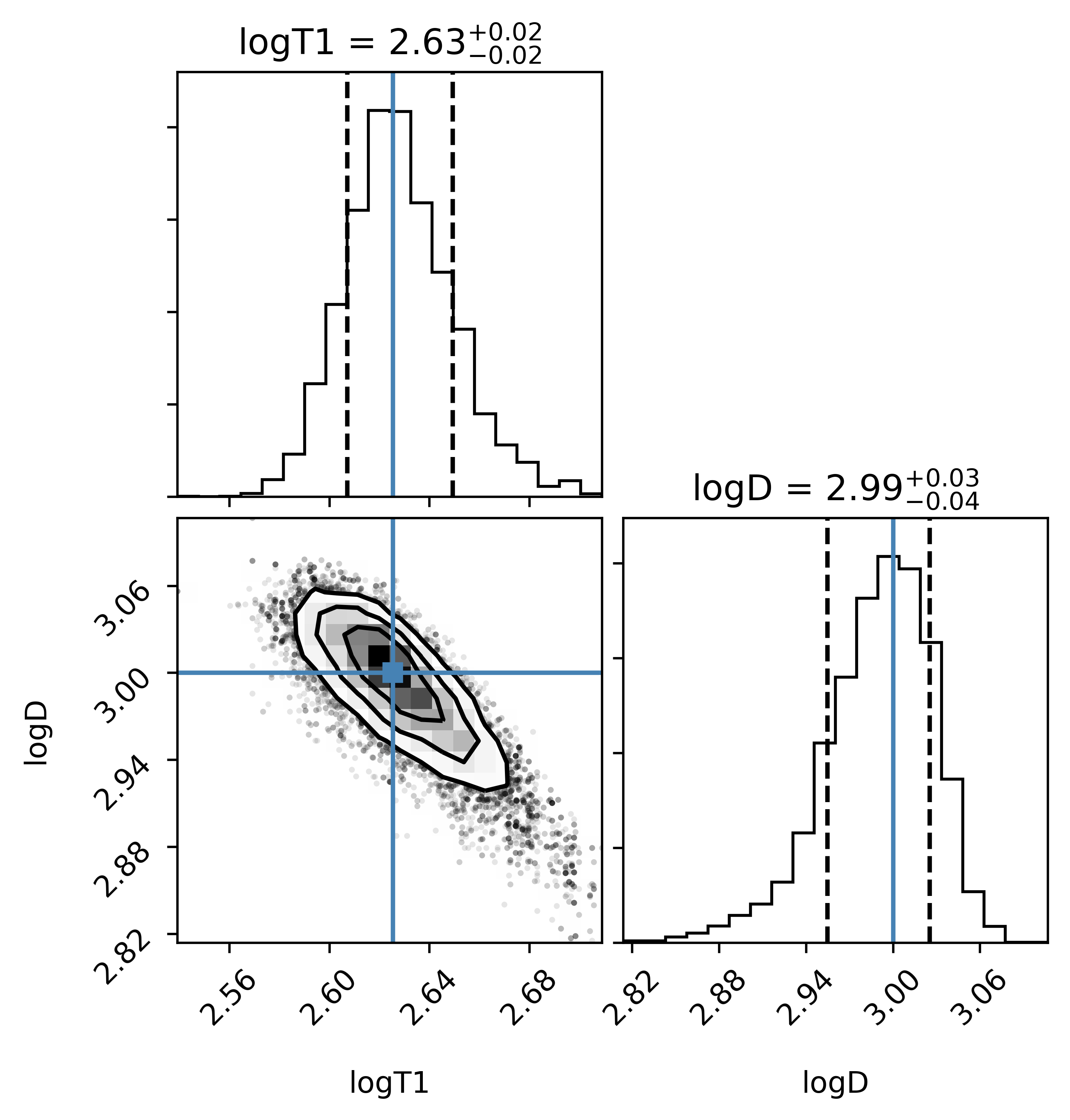}
\caption{The posterior probability density function (pdf) for the
case of fitting asteroid diameter $D$ (in m) and temperature parameter 
$T_1$ (in Kelvin; see eq.~\ref{eq:T1}). The equivalent 1$-\sigma$ 
uncertainties for $D$ and $T_1$, measured as standard deviation of
the marginal probability distributions (shown as histograms; the vertical
lines are true input values), are 8.6\% and 4.9\%, respectively. 
Note the strong covariance between $D$ and $T_1$: a smaller $D$ is 
compensated by higher $T_1$. The best-fit SED is shown in 
Fig.~\ref{fig:synthFlux}. This figure was generated using \href{https://nbviewer.jupyter.org/github/moeyensj/atm/blob/master/notebooks/validation/example_synthetic_changingNumObservations.ipynb}{notebooks/validation/example\_synthetic\_changingNumObservations.ipynb}.
\label{fig:2paramPost}}
\end{figure}

We generate synthetic fluxes in the four WISE bandpasses using an asteroid 
with $D = 1$ km, $T_1 = 422$ K, $\epsilon(\lambda)=0.7$, and asteroid-Sun 
distance of 3 A.U. The observatory is placed at 2 A.U. from the asteroid. 
The synthetic fluxes generated using the NEATM model are shown in 
Fig.~\ref{fig:synthFlux}. The four symbols with error bars correspond 
to the four in-band fluxes for the WISE observatory. As evident, the flux
in the bluest band is dominated by the reflected flux component, while the two reddest bands are dominated by the emitted flux component. When fitting for $D$ 
and $T_1$ using ``observational" constraints in the four WISE bands, we 
add Gaussian scatter to each flux with a standard deviation of 20\%. This 
scatter reflects the variability amplitude $\Sigma$ from eq.~\ref{eq:MLflux}. 

The synthetic model flux, generated using the NEATM model, is fit using 
MCMC and three different thermal models implemented in \atm: NEATM, STM
and FRM. We use 20 chains, each with 3,000 samples and 500 burn-in samples. 
The best-fit parameters obtained using the STM and FRM
models are very similar, but not identical, to the parameters obtained
with the correct NEATM model. The three best-fit SEDs for the fiducial 
validation case are shown in Fig.~\ref{fig:synthFlux}. 

The posterior probability density function (pdf) for the NEATM fitting
case is shown in Fig.~\ref{fig:2paramPost}. Note that the use of full
multi-dimensional pdf for the analysis of constraints on the model parameters
is superior to using the so-called ``point estimates'' obtained by maximum
likelihood methods (e.g., the least squares method). A strong covariance between 
$D$ and $T_1$ is clearly visible: a smaller $D$ is compensated by higher 
$T_1$, with a bias (as well as scatter) in best-fit $D$ twice as large 
as the corresponding $T_1$ bias (because the observed flux approximately 
scales with $D^2 T_1^4$). The equivalent 1$-\sigma$ uncertainties for $D$ 
and $T_1$ are 8.6\% and 4.9\%, respectively. These uncertainty levels can 
be approximately explained from first principles: given the flux uncertainty 
of 20\% and 4 data points, the overall flux normalization uncertainty is 
about 10\% (=20\%/$\sqrt{4}$). If $T_1$ were known, the best-fit uncertainty
for $D$ would be about 5\%, and if $D$ were known, the best-fit uncertainty
for $T_1$ would be about 2.5\%. Given the covariance between best-fit
$D$ and $T_1$, their expected uncertainties are about twice as large (see
Fig.~\ref{fig:2paramPost}), in good agreement with the MCMC results. 

When the number of epochs $N$ is increased, we find that the fitting uncertainty 
for the best-fit parameter decreases as $N^{-1/2}$, as expected. The fitting 
uncertainty increases approximately linearly with the assumed flux uncertainty 
($\Sigma=0.2$ mag, presumably due to variability, rather than due to measurement 
uncertainties). For example, with $\Sigma=0.1$ mag and 25 observing epochs 
(each with four WISE fluxes), the fitting uncertainty for the best-fit parameters 
$D$ and $T_1$ decreases by about a factor of ten, to 0.92\% and 0.52\%, respectively.
Of course, here we are using the same model for fitting as was used to generate 
the synthetic flux -- in reality, such small uncertainties are {\it essentially 
impossible} due to numerous systematic shortcomings of idealized thermal models 
(and uncertain $\epsilon(\lambda)$ -- see below). 

These tests assumed that the {\it correct} values of $\epsilon(\lambda)$ are 
known a priori. When we set $\epsilon_{W1W2}$ to an incorrect value of 
0.8, instead of the true input value of 0.7, we found that the best-fit
value of $D$ was biased by as much as 22\%. Therefore, {\it even though formal
fitting precision can be high, the accuracy of fitted parameters can be
significantly worse.} 

\begin{figure}[th]
\centering
\includegraphics[width=0.9\textwidth, keepaspectratio]{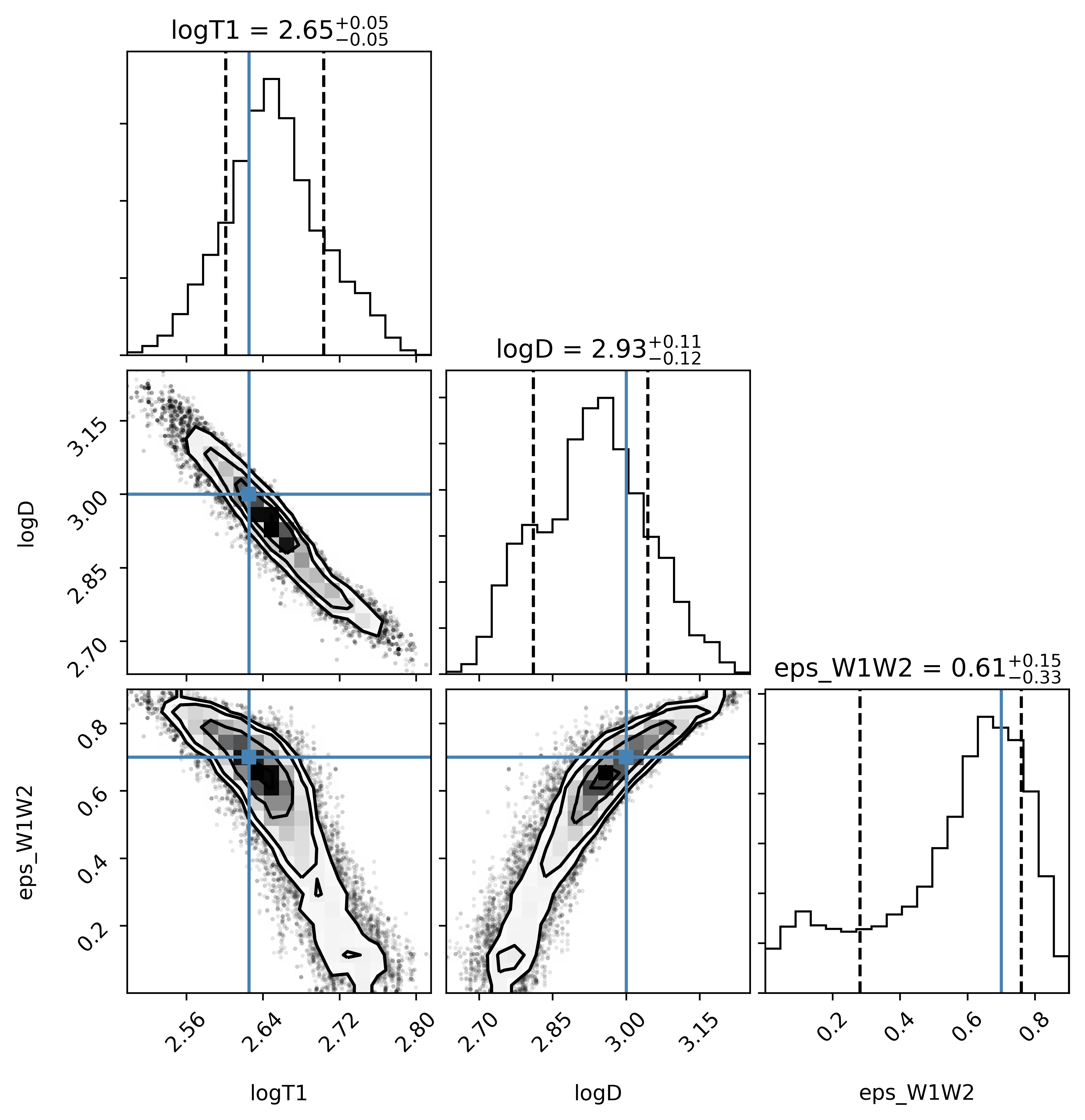}
\caption{Analogous to Fig.~\ref{fig:2paramPost}, but here  
for the case of fitting three free parameters: asteroid diameter 
$D$, temperature parameter $T_1$, and the value of $\epsilon(\lambda)$ 
in the WISE bands 1 and 2, $\epsilon_{W1W2}$. Although all other simulation 
and fitting parameters are the same as is in Fig.~\ref{fig:2paramPost}, 
the differences between the posterior pdfs are substantial. For example,
the equivalent 1$-\sigma$ uncertainties for $D$ and $T_1$ increased
from 8.6\% and 4.9\% to 30\% and 12\%, respectively. 
In addition to the covariance between $D$ and $T_1$, both 
parameters show covariances with $\epsilon_{W1W2}$. Compared
to $D$ and $T_1$, posterior constraints on $\epsilon_{W1W2}$ are
much weaker. This figure was generated using \href{https://nbviewer.jupyter.org/github/moeyensj/atm/blob/master/notebooks/validation/example_synthetic_changingNumObservations.ipynb}{notebooks/validation/example\_synthetic\_changingNumObservations.ipynb}.
\label{fig:3paramPost}}
\end{figure}

We investigated next how the precision of the best-fit $D$ and
$T_1$ decreases when emissivity $\epsilon(\lambda)$ is also fit. 
Following discussion in \S\ref{sec:eps}, we fit $\epsilon_{W1W2}$, 
the value of $\epsilon(\lambda)$ in the WISE bands 1 and 2, and
assume that $\epsilon_{W3W4}$, the value of $\epsilon(\lambda)$ in 
the WISE bands 3 and 4, is known. In other words, we assume a flat
prior in the range 0 to 1 for $\epsilon_{W1W2}$, and a Dirac 
$\delta$ function for $\epsilon_{W3W4}$, centered on its true value.
The posterior pdf is shown in Fig.~\ref{fig:3paramPost}. The differences 
compared to the posterior pdf in Fig.~\ref{fig:2paramPost} are substantial; 
for example, the equivalent 1$-\sigma$ uncertainties for $D$ and $T_1$ 
increased by about a factor of three: from 8.6\% and 4.9\% to 30\% and 
12\%, respectively. In addition to the covariance between $D$ 
and $T_1$, both parameters show covariances with $\epsilon_{W1W2}$. 
A larger $\epsilon_{W1W2}$ can be compensated by a lower $T_1$ or 
a larger $D$. For example, a 0.1 uncertainty in $\epsilon_{W1W2}$ can induce 
a $\sim$25\% uncertainty in $D$! Given that typically $\epsilon(\lambda)$ 
is not known to better than 0.05, it follows that uncertainty of best-fit 
$D$ is unlikely below $\sim$10\% (in agreement with recent analysis by \citealt{2018arXiv181101454W} and \citealt{2018AJ.156.62M}). 

\begin{figure}[th]
\centering
\includegraphics[width=0.9\textwidth, keepaspectratio]{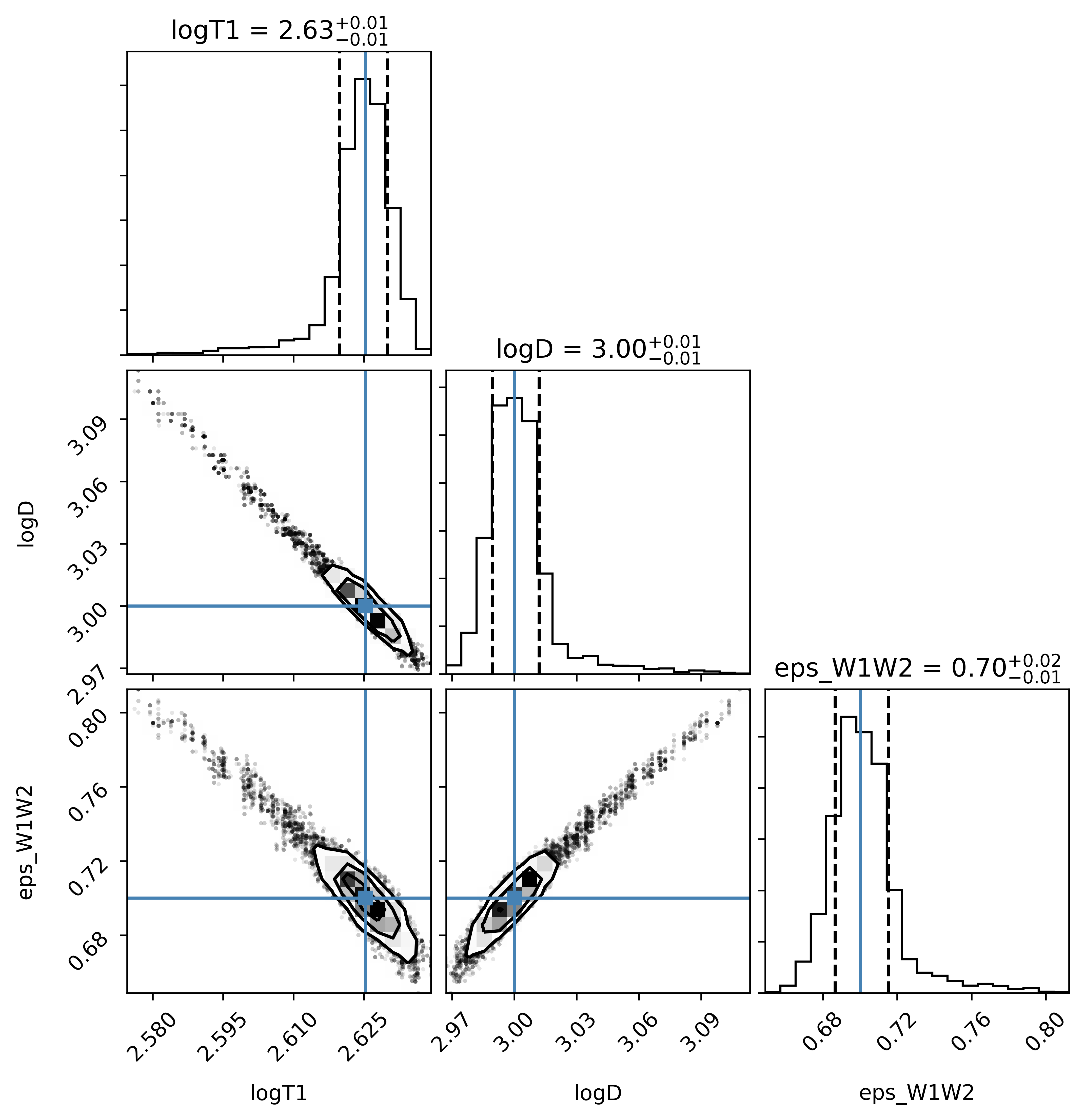}
\caption{Analogous to Fig.~\ref{fig:3paramPost}, but here with
better observational constraints: 25 epochs (each with four WISE fluxes)
instead of one, and with flux uncertainty $\Sigma$ of 0.1 mag, instead of 0.2 mag. 
Note the significant improvement in best-fit constraints, as well as
persistent covariance between the fitted parameters. This figure was generated using \href{https://nbviewer.jupyter.org/github/atm/blob/master/notebooks/validation/example_synthetic_changingNumObservations.ipynb}{notebooks/validation/example\_synthetic\_changingNumObservations.ipynb}.
\label{fig:3paramPostAcc}}
\end{figure}

Compared to $D$ and $T_1$, the posterior constraint on $\epsilon_{W1W2}$ 
is much weaker. The standard deviation for the marginal distribution of 
$\epsilon_{W1W2}$ shown in Fig.~\ref{fig:3paramPost} is 0.23, but note that 
this marginal pdf is far from Gaussian. The fitting constraints on all
three free parameters, and $\epsilon_{W1W2}$ in particular, improve as
the number of data points increases and the scatter $\Sigma$ decreases. 
Fig.~\ref{fig:3paramPostAcc} shows the posterior pdf when synthetic data
include 25 epochs (each with four WISE fluxes) instead of one in 
Fig.~\ref{fig:3paramPost}, and with flux uncertainty of $\Sigma=0.1$ mag, 
instead of 0.2 mag. As evident, the posterior constraints for all
three fitted parameters are significantly improved. The standard
deviations for the marginal distributions are 2.5\% for $D$, 1.1\%
for $T_1$ and 0.014 for $\epsilon_{W1W2}$ (recall that in the case
of fitting only $D$ and $T_1$, their uncertainties were equal to 
0.92\% and 0.52\%, respectively, using the same number of observations
and the same value of $\Sigma$). Therefore, adding $\epsilon_{W1W2}$
as a free parameter results in about twice as large statistical uncertainties for the best-fit size $D$.  

\subsubsection{Physical Validation} 

The analysis described in the preceding section validated internal code consistency 
using synthetic data. While \atm\ can reproduce its own inputs, it still needs to
be validated in an absolute sense. We now use several real asteroids, with sizes 
known independently from infrared flux measurements, to physically validate the code 
and implemented models. 
 
\begin{figure}[th]
\centering
\includegraphics[width=0.9\textwidth, keepaspectratio]{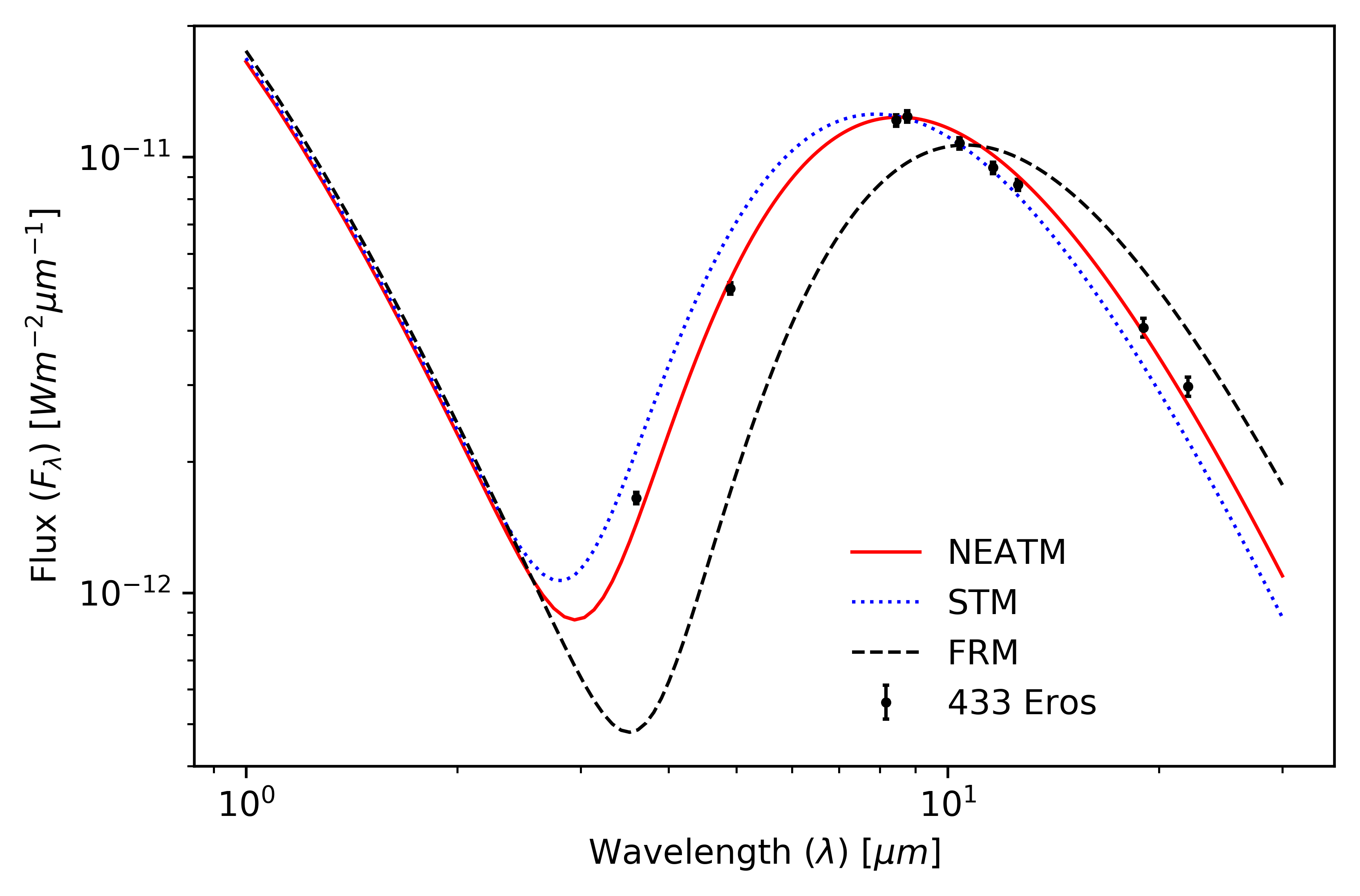}
\caption{Physical model validation using data for 433 Eros
(symbols) from \cite{1979Icar...40..297L} and models evaluated using best-fit parameters from \cite{1998Icar..131..291H}. This figure recreates Figure 1a from \cite{1998Icar..131..291H} and was generated using \href{https://nbviewer.jupyter.org/github/moeyensj/atm/blob/master/notebooks/validation/example_Eros.ipynb}{notebooks/validation/example\_Eros.ipynb}.
\label{fig:modelValid1}}
\end{figure}

\begin{figure}[th]
\centering
\includegraphics[width=0.9\textwidth, keepaspectratio]{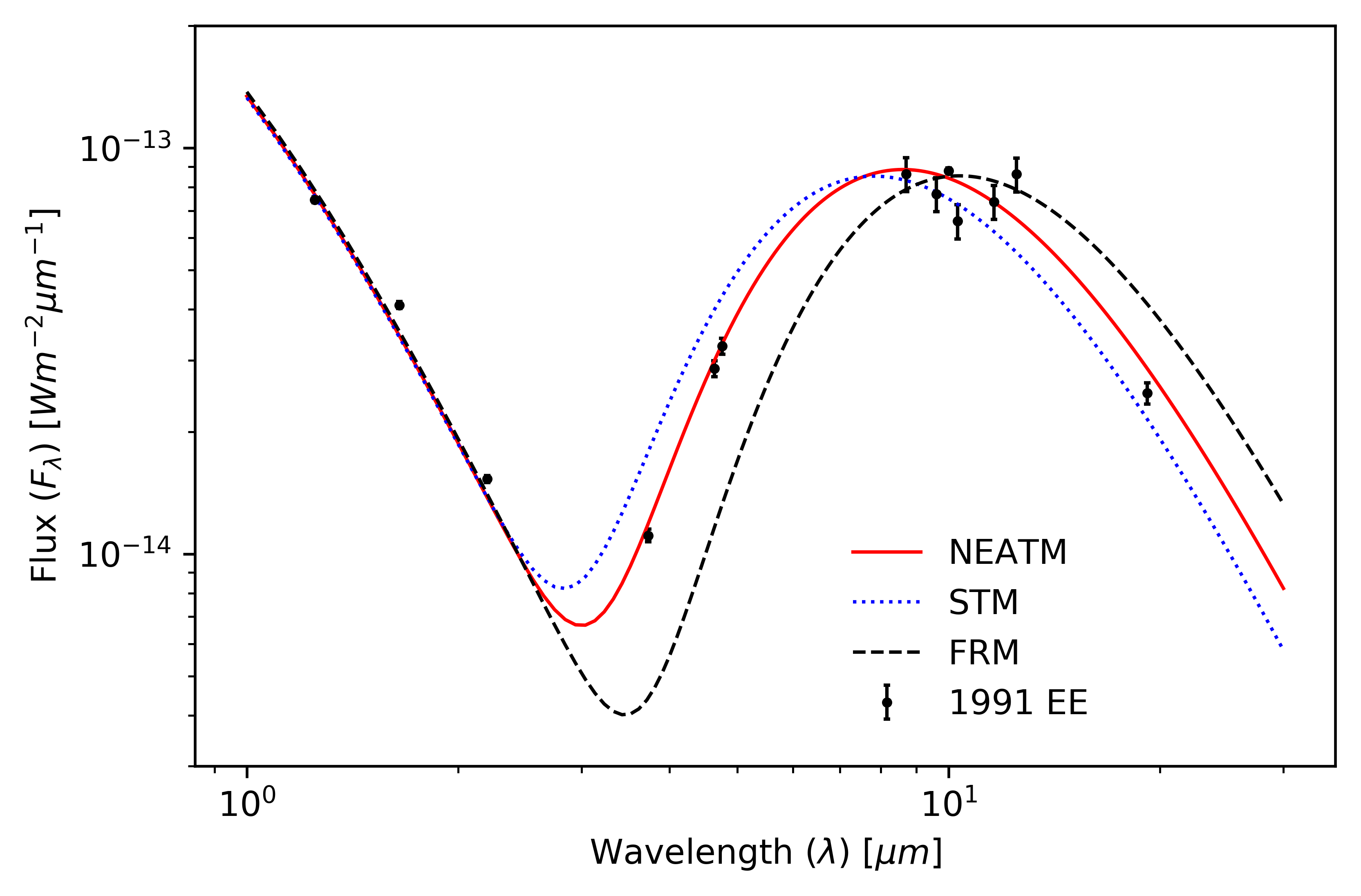}
\caption{Physical model validation using data for asteroid 1991 EE from \cite{1998Icar..135..441H}. This figure recreates Figure 3 from \cite{1998Icar..135..441H} and was generated using \href{https://nbviewer.jupyter.org/github/moeyensj/atm/blob/master/notebooks/validation/example_1991EE.ipynb}{notebooks/validation/example\_1991EE.ipynb}.
\label{fig:modelValid2}}
\end{figure}

First, we consider the observations of asteroid 433 Eros by \cite{1979Icar...40..297L}. 
We use best-fit model parameters from \cite{1998Icar..131..291H} and demonstrate in 
Fig.~\ref{fig:modelValid1} that \atm\ produces model fluxes in agreement with 
measurements. An additional example, with data and analysis for near-Earth asteroid 
1991 EE from \cite{1998Icar..135..441H}, also includes and validates the reflected 
light component, as shown in Fig.~\ref{fig:modelValid2}. Finally, \atm\ is also validated
against NEOWISE results, as described in detail in \S\ref{sec:NEOWISEATM}. 

\subsection{The Impact of Missing Bands on Best-fit Parameters}

Our validation discussion in \S\ref{sec:internalvalid} was based on synthetic data for all
four WISE bands. The WISE effective bandpass wavelengths are well 
spaced to provide simultaneous constraints on $T_1$, $D$ and 
$\epsilon_{W1W2}$, given a sufficient number of observations. 
We have shown above that fitting precision deteriorates as $\epsilon_{W1W2}$
is added as a fitting parameter: with 25 epochs and $\Sigma=0.1$ mag,  the uncertainty 
of best-fit $D$ and $T_1$ increases by about a factor of two from 0.92\% and 0.52\% to 
2.5\% and 1.1\%, respectively (with an uncertainty of 0.012 for $\epsilon_{W1W2}$). 
Again, these are formal fitting uncertainties when the model is perfectly well known. 
We now investigate how the precision further deteriorates when only a subset of 
WISE bandpasses is used in fitting.

When only $T_1$ and $D$ are fit, the largest increase in uncertainties, about a factor of 
two, is seen when the missing bandpass is $W_1$. This is because the total flux in the 
$W_1$ bandpass is dominated by the reflected sunlight component and thus it depends very 
weakly on $T_1$. Consequently, $W_1$ strongly constrains diameter $D$ in the case of known 
$\epsilon$ (and thus known albedo). Indeed, when only the $W_1$ and $W_2$ 
bandpasses are used, the $T_1$ precision worsens by only 50\% and the $D$ precision
remains unchanged. When only $W_1$ and $W_3$ are used, there is no appreciable decrease in precision. 
On the other hand, a combination of the $W_2$ and $W_3$ bandpasses results in 3 to 4 times 
larger uncertainties. Somewhat surprisingly, the fitter does not recover the correct 
parameters when only the $W_3$ and $W_4$ bandpasses are used. In the latter case,
typically the best-fit temperature is under-estimated, while the best-fit size 
can be larger than the input value by close to a factor of 10.  
When fitting three parameters ($T_1$, $D$ and $\epsilon_{W1W2}$), flux in at least 
one of the two bluest bandpasses must be available to recover the input parameters.  
A decrease in precision of less than a factor of two is observed when fitting with three bandpasses 
as long as the $W_4$ band is used in fitting. When the $W_4$ data are not used, 
the fitting precision decreases by about a factor of three compared to the 
four-bandpass case. 

In summary, when only $T_1$ and $D$ are fit, even two bandpasses will suffice as long
as $W_1$ or $W_2$ is included, but when $\epsilon_{W1W2}$ is also fit, $W_4$ 
becomes more important than $W_1$ and $W_2$. However, note the caveat that these conclusions are
derived for the somewhat unrealistic case with a lot of observations and small 
$\Sigma$.

\section{Illustration of ATM fitting capabilities\label{sec:use}}

\begin{figure}[th]
\centering
\includegraphics[width=0.9\textwidth, keepaspectratio]{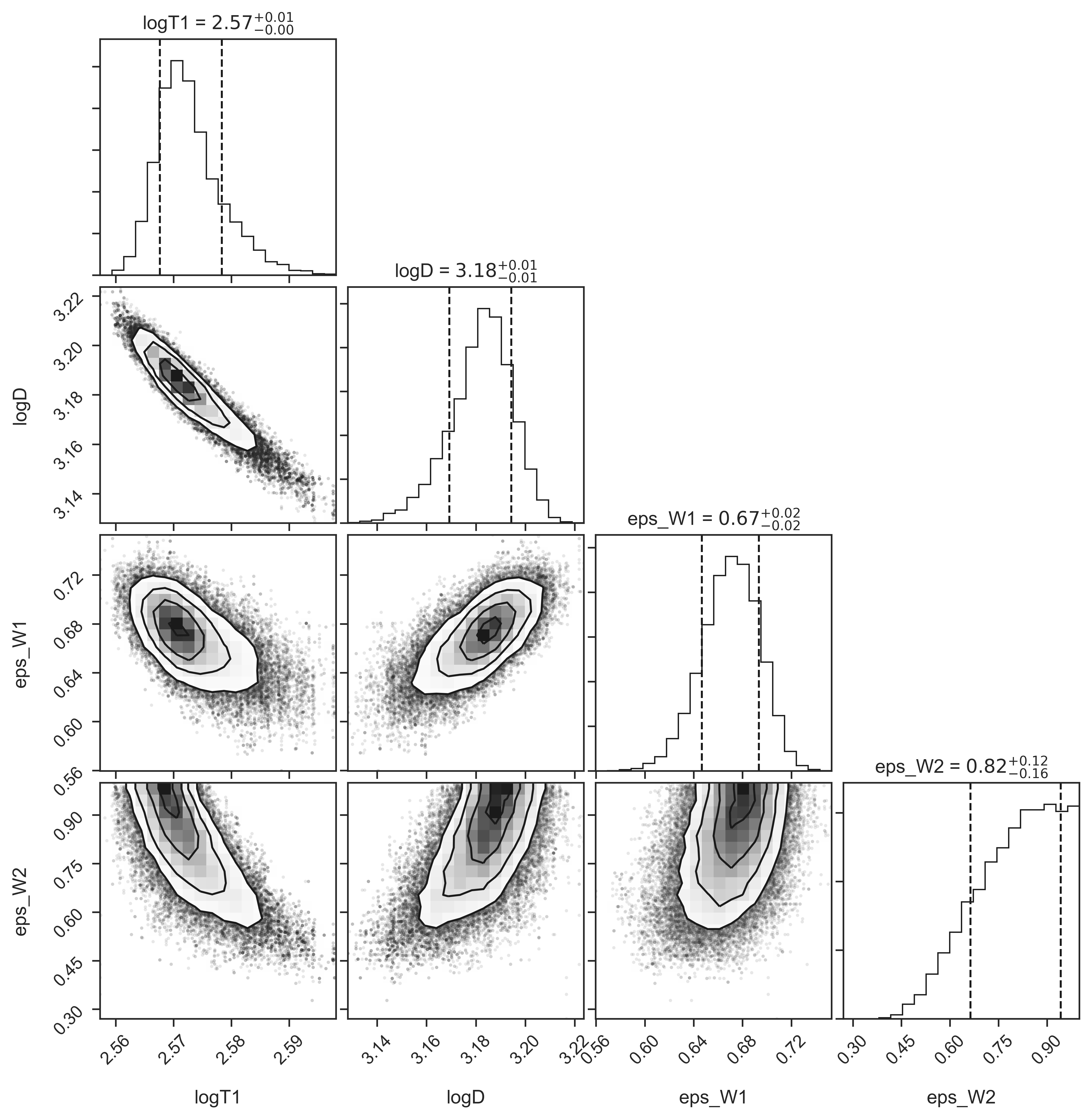}
\caption{\atm\ ``corner'' plot for asteroid (54789) assuming model 5. Here
both $\epsilon_{W1}$ and $\epsilon_{W2}$ are free parameters; compare to 
Figure~\ref{fig:corner54789model4} where an additional constraint 
$\epsilon_{W1} = \epsilon_{W2}$ is imposed. This figure was generated using \href{https://nbviewer.jupyter.org/github/moeyensj/atm/blob/master/notebooks/analysis/single_object_54789.ipynb}{notebooks/analysis/single\_object\_54789.ipynb}.
\label{fig:corner54789model5}}
\end{figure}

\begin{figure}[th]
\centering
\includegraphics[width=0.9\textwidth, keepaspectratio]{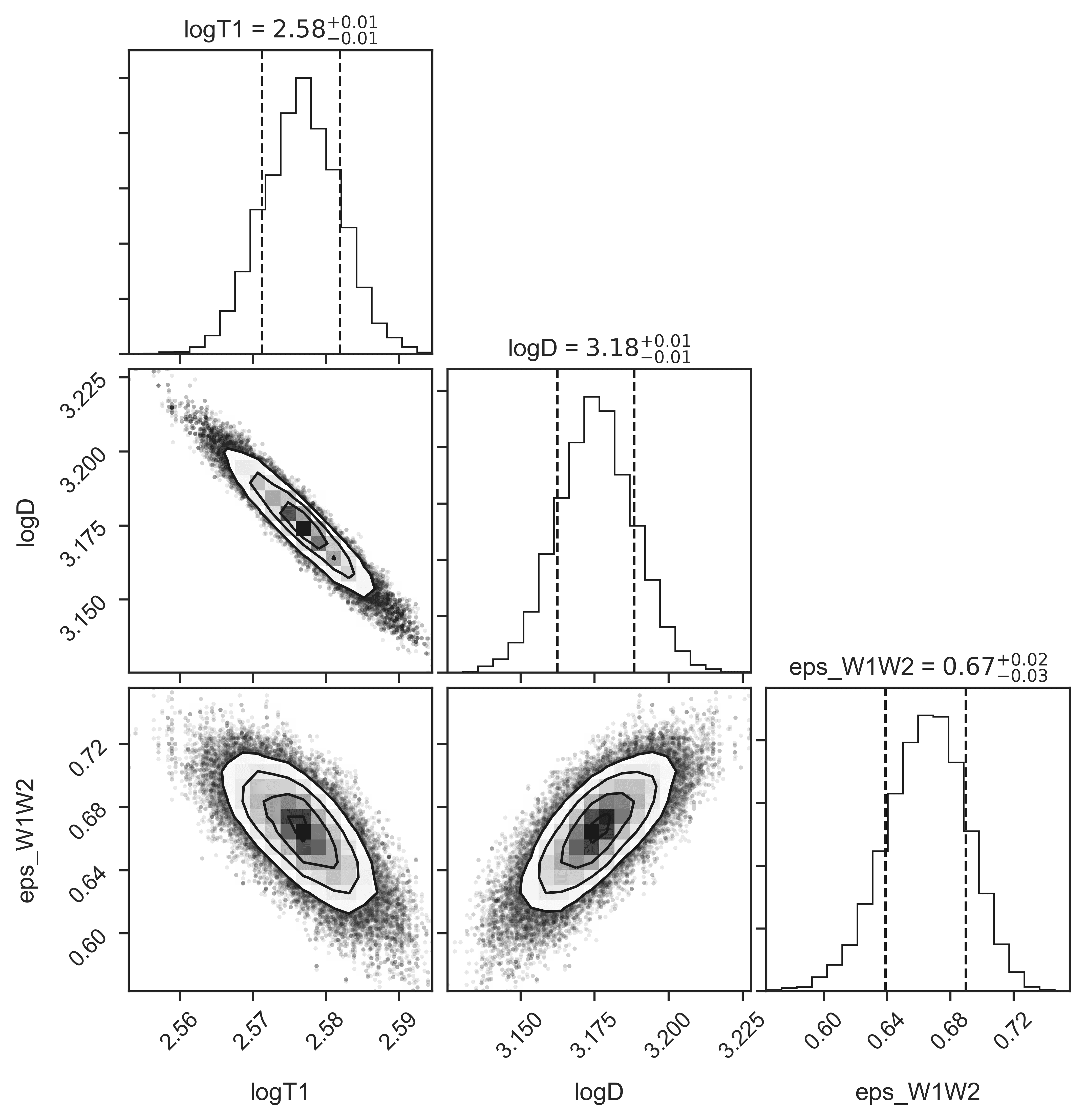}
\caption{\atm\ ``corner'' plot for asteroid (54789) assuming model 4. This
model imposes a constraint  $\epsilon_{W1} = \epsilon_{W2}$. Compare to 
Figure~\ref{fig:corner54789model5} where both $\epsilon_{W1}$ and 
$\epsilon_{W2}$ are free parameters. This figure was generated using \href{https://nbviewer.jupyter.org/github/moeyensj/atm/blob/master/notebooks/analysis/single_object_54789.ipynb}{notebooks/analysis/single\_object\_54789.ipynb}.
\label{fig:corner54789model4}}
\end{figure}

In this section, we illustrate \atm's capabilities for fitting asteroid spectral energy distributions, 
with emphasis on the treatment of Bayesian priors when also fitting emissivity, using three 
well-observed asteroids: (25916), (54789), and (90367). We study how the choices of priors affect the posterior pdf for fitted parameters and we study the 
resulting biases in point estimates derived from these pdfs (that is, ``best-fit parameters''). 
We consider five different models, where in addition to fitting for asteroid diameter, $D$,
and temperature parameter, $T_1$, we treat emissivity $\epsilon(\lambda)$ as follows:
\begin{itemize}
\item {\bf Model 1:} $\epsilon(\lambda) = \epsilon_0 =$ const. and we fit for $\epsilon_0$ using 
       a flat prior $0 < \epsilon_0 < 1$.
\item {\bf Model 2:}  We set $\epsilon_{W3} = \epsilon_{W4} = 0.9$ (that is, the prior is the Dirac $\delta$ function) and fit 
for unknown $\epsilon_{W1} = \epsilon_{W2} = \epsilon_{W1W2}$, where subscripts indicate WISE bands.
Here, and in models below, we model $\epsilon(\lambda)$ as a step function, with 
transition wavelengths\footnote{See \url{http://www.astro.ucla.edu/~wright/WISE/passbands.html}}
between bands at $\lambda_{W1W2} = 3.9 \, \mu{\rm m}$, $\lambda_{W2W3} = 6.5 \, \mu{\rm m}$,  and 
$\lambda_{W3W4} = 18.5 \, \mu{\rm m}$. Note that in SED plots discussed below emissivity $\epsilon_{W1}$  
is extrapolated to wavelengths shorter than $\lambda_{W1W2}$, implying a pre-set value for the visual
albedo, $p_V$. The same is true for emissivity in the most redward band, $\epsilon_{W4}$; it is extrapolated to 
wavelengths beyond the W4 bandpass for plotting purposes. 
\item {\bf Model 3:}  We set $\epsilon_{W3} = 0.70$ and $\epsilon_{W4} = 0.86$, motivated by ensemble 
analysis presented later in \S\ref{sec:FitAnalysis2}, and fit for $\epsilon_{W1W2}$.
\item {\bf Model 4:}  We set $\epsilon_{W3} = 0.80$ and $\epsilon_{W4} = 0.98$ and fit for $\epsilon_{W1W2}$.
The $\epsilon_{W4}/\epsilon_{W3}$ ratio is the same as in the previous model. 
\item {\bf Model 5:}  We set $\epsilon_{W3} = 0.80$ and $\epsilon_{W4} = 0.98$ and fit for $\epsilon_{W1}$
and $\epsilon_{W2}$. Compared to the previous model, here we do not enforce $\epsilon_{W1} 
= \epsilon_{W2}$ . 
\end{itemize}
 
Figure~\ref{fig:corner54789model5} shows the posterior pdf for four fitted parameters in the case
of model 5 and asteroid (54789). We have also considered a case where emissivity values in all 
four bands are free fitting parameters, for a total of six fitted parameters, but concluded that it 
is sufficient to start the discussion here with model 5. As panels in the bottom row of 
Figure~\ref{fig:corner54789model5} show, the pdf for $\epsilon_{W2}$ is quite wide and not too dissimilar
to its prior, while $D$, $T_1$ and $\epsilon_{W1}$ have pdfs that are much narrower than their priors, 
and with a well defined peak. In other words, $\epsilon_{W2}$ is much less constrained by the data 
than the other three parameters. This conclusion is also valid for the other two asteroids considered 
here. 
 
Following the behavior of emissivity inferred from laboratory spectra for different materials, as
illustrated in Figure 1 from \cite{NM2018a}, we enforce an additional fitting constraint: 
$\epsilon_{W1} = \epsilon_{W2}= \epsilon_{W1W2}$ and fit for three free parameters (models 2, 3, and 4).
Figure~\ref{fig:corner54789model4} shows the posterior pdf for fitted parameters in the model 4 case. 
The fitted parameter $\epsilon_{W1W2}$ now has a pdf that is much narrower than its
priors and with a well defined peak. The same behavior is observed for the other two asteroids.
As a result of this analysis, we conclude that for the robust fitting of these, and other less observed,
asteroids, we need to only fit for $\epsilon_{W1W2}$, and not for $\epsilon_{W1}$ and $\epsilon_{W2}$ 
separately. 

There is a remaining question of what to choose for the values of $\epsilon_{W3}$ and $\epsilon_{W4}$ 
that are not fitting parameters. As Table~\ref{tab:SEDs} shows, models 2, 3, and 4 result in
values of best-fit $D$ varying by about 10-15\%, depending on the chosen values of $\epsilon_{W3}$ 
and $\epsilon_{W4}$. In addition, adopting $\epsilon(\lambda) = \epsilon_0$ and fitting for $\epsilon_0$ 
can result in a change of $D$ as large as 20\% (compare model 1 and model 2 for asteroid 54789). 
This approach also leads to a taxonomy-dependent $D$ bias (systematic uncertainty) because the 
actual bias depends on detailed deviations of $\epsilon(\lambda)$ from the assumed constant value of $\epsilon_0$. 
We will return to the discussion of optimal choice of $\epsilon_{W3}$ and $\epsilon_{W4}$ in 
\S\ref{sec:FitAnalysis2}.

\begin{table} 
\begin{center} 
\caption{The Best-fit NEATM Parameters$^a$ for Asteroids (25916), (54789) and (90367)} 
\label{tab:SEDs} 
\resizebox{0.7\linewidth}{!}{
\begin{tabular*}{9.5cm}{lcccccc} 
\hline 
\hline 
Model &  log($D$/m)  &   log($T_1$/K)  &  $\epsilon_{W1}$ &  $\epsilon_{W2}$ &  $\epsilon_{W3}$ &  $\epsilon_{W4}$ \\
\hline
\multicolumn{7}{l}{\it Asteroid (25916)} \\ 
   1    &     3.691    &       2.587     &     0.877     & =$\epsilon_{W1}$ & =$\epsilon_{W1}$ & =$\epsilon_{W1}$  \\ 
   2    &     3.681    &       2.589     &     0.874     & =$\epsilon_{W1}$ &     (0.90)    &     (0.90)     \\ 
   3    &     3.702    &       2.585     &     0.884     & =$\epsilon_{W1}$ &     (0.70)    &     (0.86)     \\  
   4    &     3.684    &       2.589     &     0.877     & =$\epsilon_{W1}$ &     (0.80)    &     (0.98)     \\  
   5    &     3.664    &       2.602     &     0.902     &     0.660        &     (0.80)    &     (0.98)     \\  
\hline
\multicolumn{7}{l}{\it Asteroid (54789)} \\ 
   1    &     3.240    &       2.558     &     0.740     & =$\epsilon_{W1}$ & =$\epsilon_{W1}$ & =$\epsilon_{W1}$  \\ 
   2    &     3.166    &       2.575     &     0.653     & =$\epsilon_{W1}$ &     (0.90)    &     (0.90)     \\ 
   3    &     3.198    &       2.571     &     0.697     & =$\epsilon_{W1}$ &     (0.70)    &     (0.86)     \\  
   4    &     3.175    &       2.577     &     0.665     & =$\epsilon_{W1}$ &     (0.80)    &     (0.98)     \\  
   5    &     3.184    &       2.572     &     0.672     &     0.827        &     (0.80)    &     (0.98)     \\  
\hline
\multicolumn{7}{l}{\it Asteroid (90367)} \\ 
   1    &     3.223    &       2.580     &     0.971     & =$\epsilon_{W1}$ & =$\epsilon_{W1}$ & =$\epsilon_{W1}$  \\ 
   2    &     3.254    &       2.576     &     0.972     & =$\epsilon_{W1}$ &     (0.90)    &     (0.90)     \\ 
   3    &     3.278    &       2.572     &     0.974     & =$\epsilon_{W1}$ &     (0.70)    &     (0.86)     \\  
   4    &     3.261    &       2.575     &     0.974     & =$\epsilon_{W1}$ &     (0.80)    &     (0.98)     \\  
   5    &     3.260    &       2.576     &     0.976     &     0.953        &     (0.80)    &     (0.98)     \\  
\hline
\multicolumn{7}{l}{(a) The values in parenthesis represent priors.} \\
\end{tabular*} 
}
\end{center} 
\end{table} 

\begin{figure}[th]
\centering
\includegraphics[width=0.9\textwidth, keepaspectratio]{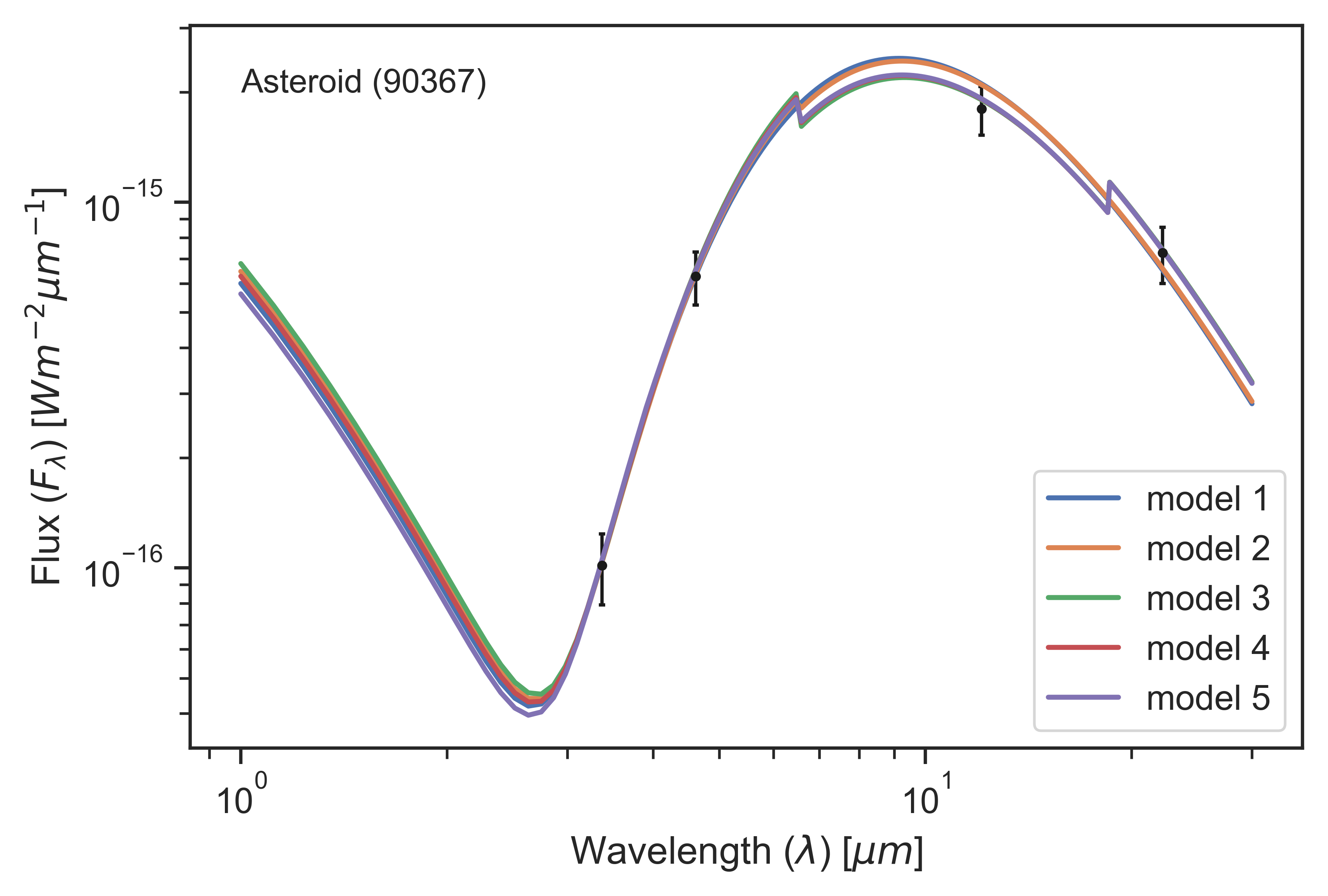}
\caption{The symbols show the median fluxes and their errors for 
WISE data for asteroid (90367). The errors include the contribution of
assumed variability amplitude of 0.15 mag. The solids line show five 
NEATM models that differ in chosen priors for emissivity (see text). The best-fit
parameters for these models are listed in Table~\ref{tab:SEDs}.  The 
sharp changes in modeled fluxes at $\lambda_{W2W3} = 6.5 \, \mu{\rm m}$ and 
$\lambda_{W3W4} = 18.5 \, \mu{\rm m}$ is due to a simplistic approximation 
of emissivity as a function of wavelength as a step function. This figure was generated using \href{https://nbviewer.jupyter.org/github/moeyensj/atm/blob/master/notebooks/analysis/single_object_90367.ipynb}{notebooks/analysis/single\_object\_90367.ipynb}.
\label{fig:SED90367}} 
\end{figure}

Figure~\ref{fig:SED90367} compares spectral energy distributions for models 1-5
in the case of asteroid (90367). As evident, all models are equally successful in 
explaining observed fluxes in bands W1 and W2. Models 3, 4 and 5 are more 
successful than models 1 and 2 in explaining observed fluxes in bands W3 and W4. 
This improvement is due to the choice $\epsilon_{W4} =  1.22 \epsilon_{W3}$, instead
of $\epsilon_{W4} =  \epsilon_{W3}$ for models 1 and 2, and is discussed in detail 
in \S\ref{sec:FitAnalysis2}.  This model degeneracy implies a bias in best-fit 
asteroid size in the range 10-20\%, as discussed above. 

Note that the prediction for reflected flux at wavelengths
below $\sim 2 \mu{\rm m}$ varies by about 30\% due to variation of best-fit $D$ and
best-fit $\epsilon_{W1}$, which implies an extrapolated value of $p_V$. Nevertheless, 
a measurement of flux at optical wavelengths could only break the model degeneracy 
if there were a strong prior reason to believe that emissivity values at optical 
wavelengths were somehow fully determined by the value of emissivity in the WISE W1 band.

\section{Application of ATM to WISE data and comparison with NEOWISE analysis\label{sec:NEOWISEATM}}

In this section, we apply \atm\ to a ``gold'' sample of $\sim$7,000 best-observed asteroids by WISE.
In addition to the best-fit diameter $D$ and characteristic temperature $T_1$ for each object, 
we also obtain best-fit emissivity across bands W1 and W2, $\epsilon_{W1W2}$. We compare 
\atm\ best-fit parameters to their values as published by the NEOWISE team, discuss physical implications
of the best-fit parameters, and derive an approximate method for estimating $D$ from WISE W3
measurements that doesn't require model fitting and is applicable to the majority of asteroids 
with WISE measurements. 

\subsection{Selection of High-quality WISE Data and Reliable \atm\ Fits \label{sec:WISEdtsel}} 

To cleanly compare \atm\ best-fit parameters with their values as published by the NEOWISE
team, we select a relatively small subsample ($\sim5$\% of the full sample) with the highest quality
and quantity of WISE data. Observations of asteroids by WISE were obtained using the same criteria described 
by NEOWISE papers.  The list of observations were obtained from the Minor Planet Center database for 
observatory code C51.   The positions and times were then used to search the WISE All-Sky Single Exposure 
Level 1b Source catalog using a cone search with radius 10 arcseconds, and an observation time tolerance of 
$\pm 2$ seconds.  The distance from the asteroid to the Sun, distance from the asteroid to WISE, and
the phase angle were obtained with the JPL HORIZONS service\footnote{See \url{https://ssd.jpl.nasa.gov/?horizons}}, 
using the asteroid designation and observation 
time. To ensure high quality measurements, we require\footnote{For a detailed discussion of WISE data products see \url{http://wise2.ipac.caltech.edu/docs/release/prelim/expsup/sec2\_2a.html}}:
\begin{itemize}
\item at least 3 observations with the photometric signal-to-noise ratio of at least 4 in each band, and
\item artifact flag = 0 and quality flag = A, B, C, but not U or X. 
\end{itemize}
This set of criteria reduces $\sim2.5$ million observations of $\sim$150,000 asteroids to $\sim$87,000 
observations of 9,672 asteroids (each observation with fluxes measured in up to four WISE bands).  

For some objects, measurements still include outliers despite the above quality cuts. We clip outliers 
using a simple automated iterative algorithm. We first reject all points that are more than 1 mag away 
from the median magnitude (per band), then fit a straight line to magnitude vs. time relation in each 
band using the surviving data points, and finally reject all points more than 1 mag away from the best fit line. 
This outlier clipping reduced the aforementioned 9,672 asteroids down to 7,363 ($\sim$14\% of observations at this stage were
cut as magnitude outliers). Finally, 4 objects were removed as they had missing assumptions on physical parameter, $G$. The final ``gold'' sample contains $\sim$ 77,000 observations of 7,359 objects.\footnote{See \url{https://github.com/moeyensj/atm/blob/master/data/sample.db}}

For each object, we fit for diameter $D$, characteristic temperature $T_1$ and emissivity 
across the W1--W2 wavelength range, $\epsilon_{W1W2}$. We produced four sets of fits that
differ in priors for emissivity across the W3--W4 wavelength range. Following NEOWISE 
analysis, we adopted $\epsilon_{W3}=\epsilon_{W4}=0.90$ for the first set. Informed by discrepancies
between the observed and modeled distributions of objects in the W3-W4 vs. W2-W3 color-color
diagram (see below for detailed discussion), for the other three sets we adopted 
$(\epsilon_{W3}, \epsilon_{W4})$ = (0.80, 0.98), (0.76, 0.93) and (0.70, 0.86). 

\begin{figure}[th]
\centering
\includegraphics[width=0.9\textwidth, keepaspectratio]{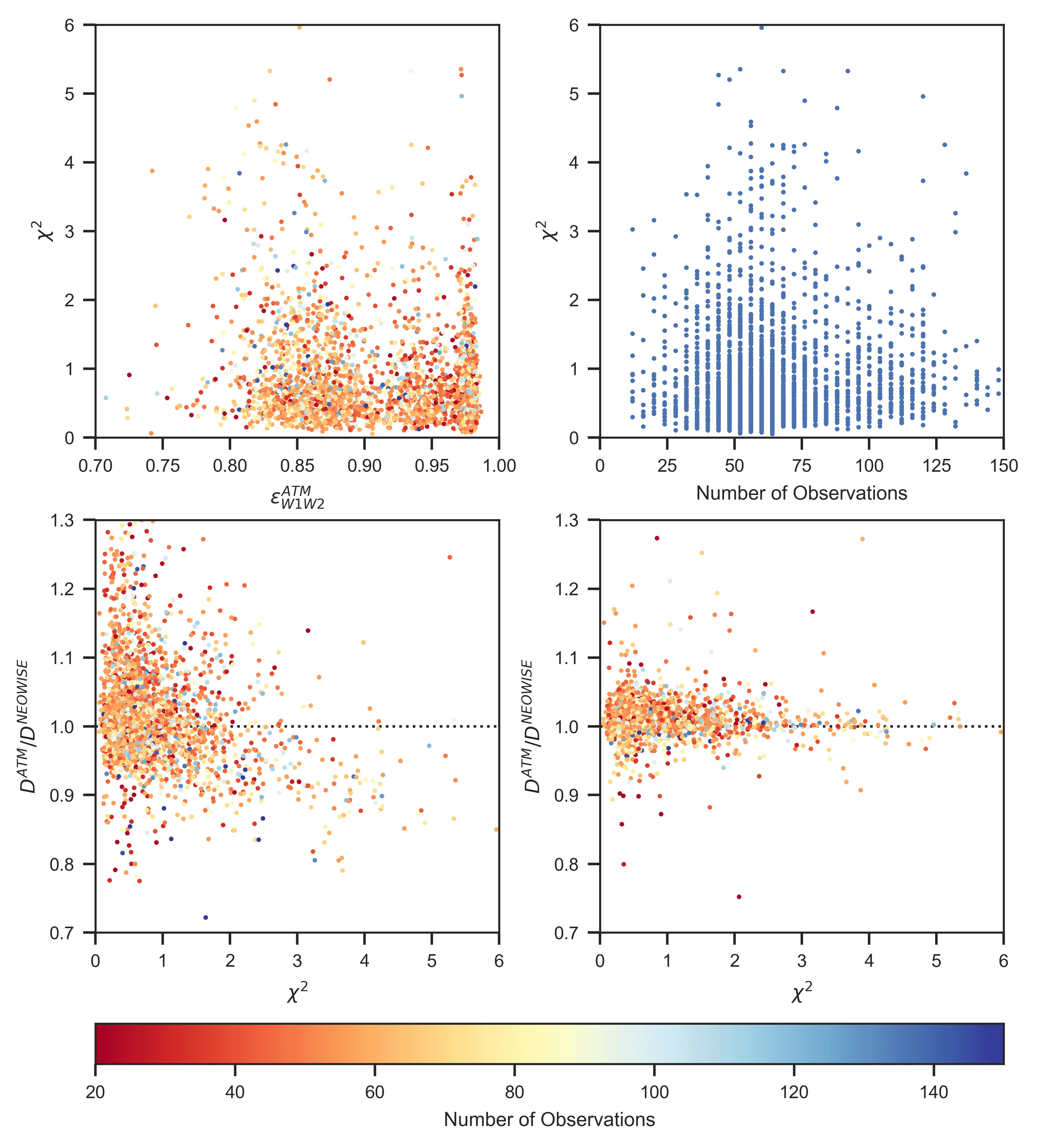}
\caption{The top left panel shows the $\chi^2$ per degree of freedom vs. $\epsilon_{W1W2}$ diagram
for asteroids that pass initial quality cuts (see \S\ref{sec:WISEdtsel}), have NEOWISE fit code 
``DVBI'' and have at least 3 observations in each of the four WISE bands. The best-fit $\chi^2$ 
and emissivity $\epsilon_{W1W2}$ were obtained using \atm. The symbols
are color-coded using the total number of data points, according to 
the legend below the panel. The top right panel shows the $\chi^2$ vs. 
the total number of data points diagram. The bottom panels show
the ratios of the best-fit diameter (left) and $\epsilon_{W1W2}$ 
(right) obtained by \atm\ and the 2016 NEOWISE release values vs. $\chi^2$,
color-coded the same way as in the top left panel. This figure was generated using \href{https://nbviewer.jupyter.org/github/moeyensj/atm/blob/master/notebooks/analysis/analysis.ipynb}{notebooks/analysis/analysis.ipynb}.
\label{fig:define1}}
\end{figure}

The top right panel in Figure~\ref{fig:define1} shows the distribution of $\chi^2$ per degree of 
freedom as a function of the total number of data points in all 4 bands. The computation of 
$\chi^2$, as well as fitting, assumes an intrinsic scatter due to variability of $\Sigma=0.15$
mag (see eq.~\ref{eq:MLflux}). The distribution of $\chi^2$, with a mode at unity, validates 
this choice. There is no correlation of the $\chi^2$ distribution with the number of data
points; typically, there are $\sim$50 measurements per object. 

\subsection{Comparison to NEOWISE Fits \label{sec:FitAnalysis1}} 

To compare our best-fit parameters to those obtained by the NEOWISE team,
we match objects from our dataset to objects from the 2016 Planetary Data System
version of the NEOWISE Diameters and Albedos 
database\footnote{See \url{https://sbn.psi.edu/pds/resource/neowisediam.html}}
\citep{2016PDSS..247.....M}, hereafter ``NEOWISE values''. For this initial comparison,
we also require NEOWISE fit code ``DVBI'' and at least 3 observations in each of the
four WISE bands, yielding a comparison sample of 2,807 asteroids. We selected only 
the NEOWISE results for each asteroid that used the most data points as some asteroids had 
multiple fitting results under the same fit code. 
The ratio of best-fit \atm\ and NEOWISE values vs. $\chi^2$ for diameter $D$ 
and emissivity $\epsilon_{W1W2}$ is shown in the bottom panels in Figure~\ref{fig:define1}. 
As evident, there is a good agreement between the two sets of best-fit parameters.

\begin{figure}[th]
\centering
\includegraphics[width=0.9\textwidth, keepaspectratio]{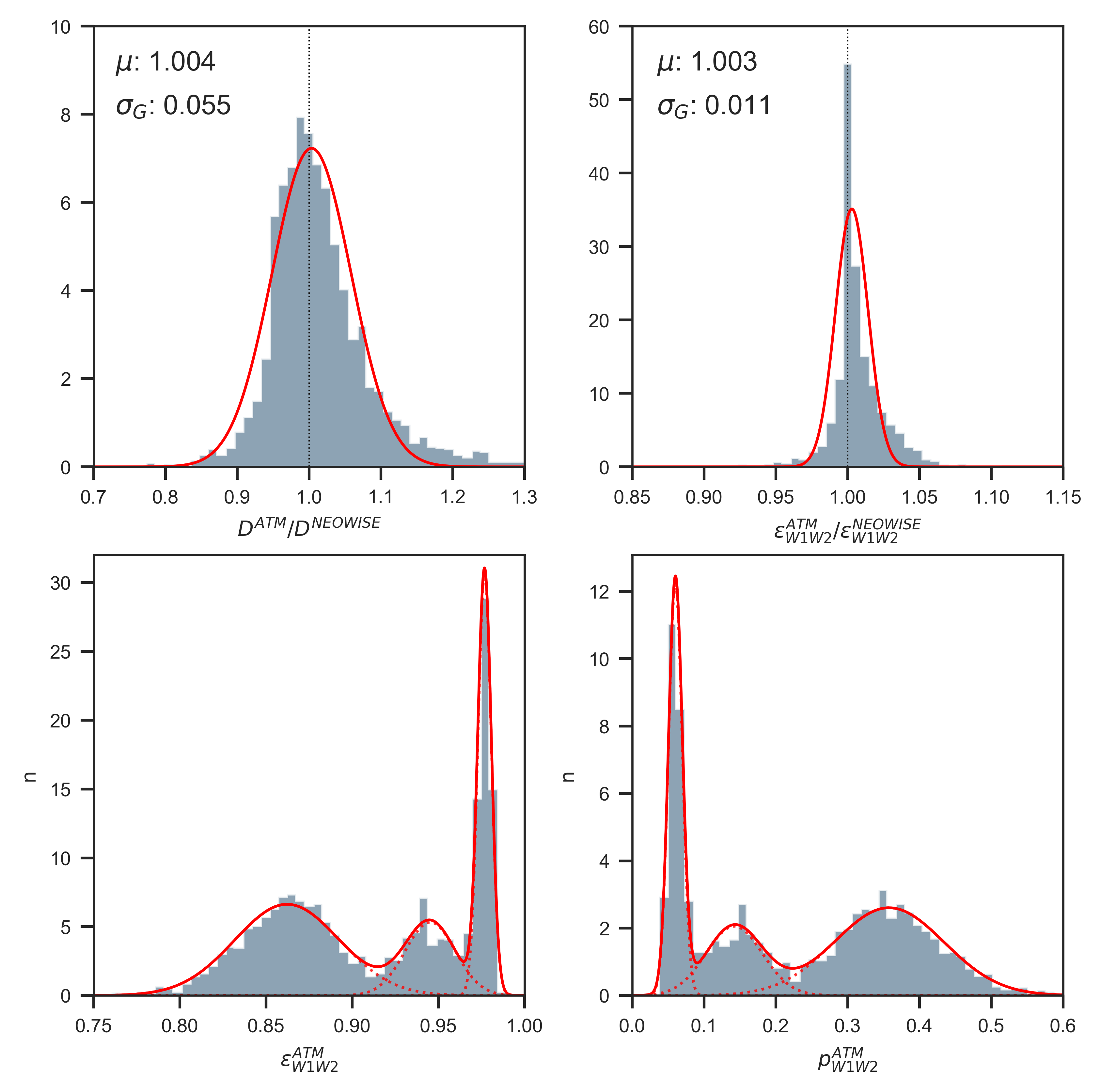}
\caption{The top panels show histograms of the ratios of the best-fit 
diameter (left) and $\epsilon_{W1W2}$ (right) \atm\ and the 2016 
NEOWISE release values, for 2,656 asteroids with $\chi^2 < 3$
(\atm) and at least 28 observations. The red lines are single 
Gaussian fits, with their mean and standard deviation shown in
each panel (e.g., \atm\ matches NEOWISE fits with a bias of
0.3\% and a scatter of 6.5\%, though note the distribution is 
leptokurtic). The bottom panels show distributions of emissivity 
$\epsilon_{W1W2}$ (left) and corresponding albedo (right) for
the same objects. The solid lines show best-fit 3-component 
Gaussian mixtures (fit to individual data points rather than to 
histograms which are shown only for illustration). The best-fit
parameters are listed in Table~\ref{tab:albedoGaussians}. This figure was generated using \href{https://nbviewer.jupyter.org/github/moeyensj/atm/blob/master/notebooks/analysis/analysis.ipynb}{notebooks/analysis/analysis.ipynb}.
\label{fig:define2}}
\end{figure}

A more quantitative comparison is shown in the top two panels in Figure~\ref{fig:define2},
for the fitting case with $\epsilon_{W3} = \epsilon_{W4}$ = 0.90. Both the best-fit diameters 
and emissivity agree on average to within 0.4\%, with a scatter of 5.5\% for diameters and 1.1\% 
for emissivity. Plausible reasons for the scatter include different outlier rejection algorithms, 
different treatments of Kirchhoff's law\footnote{We have run ATM with a fiducial dataset
both ways: with Kirchhoff's law properly implemented and with the NEOWISE ansatz. The 
distribution of the ratio of best-fit sizes is centered on 1.003, with a scatter of 1.8\%. 
In other words, although neglecting Kirchhoff's law is physically wrong, the resulting 
approximation does not contribute significantly to systematic size uncertainty when
considering WISE dataset.} (for a detailed discussion, see \citealt{NM2018a}), and how
\atm\ accounts for intrinsic variability, although we cannot exclude other causes (e.g., 
a slight error in the quadrature formula for the W3 band, see Appendix A for details). 
Whatever the reason, the discrepancies are encouragingly small. 

Our \atm\ results faithfully match the tri-modal distribution of $\epsilon_{W1W2}$ emissivity 
discovered by \cite{2014ApJ...791..121M}. The two bottom panels in Figure~\ref{fig:define2} 
show the resulting distributions of emissivity and corresponding albedo based on \atm\ results. 
The parameters of a best-fit 3-component Gaussian mixture model are listed in Table~\ref{tab:albedoGaussians}. 
Note that the fractions of three components listed in Table~\ref{tab:albedoGaussians} (20\%, 30\% and 50\%) 
are not corrected for the various sample selection effects. 

This tri-modality is related to taxonomic classes; according to \cite{2014ApJ...791..121M}
the low and high p$_{W1W2}$ albedo peaks correspond to the low and high visible albedo groups
observed previously (C/D/P and S groups, respectively), while the intermediate albedo peak 
corresponds to intermediate visible albedo values that are blended with the high albedo objects
in visible albedo distributions.


%

%

\begin{deluxetable}{l|r|r|r|r|r|r|r|r|r|}[t]
\tablecaption{The Best-fit Parameters for Emissivity and Albedo Distributions\tablenotemark{a} \label{tab:albedoGaussians}}
\tablehead{
\colhead{Quantity} & \colhead{fraction$_1$} & \colhead{$\mu_1$} & \colhead{$\sigma_1$} &
\colhead{fraction$_2$} & \colhead{$\mu_2$} & \colhead{$\sigma_2$} &
\colhead{fraction$_3$} & \colhead{$\mu_3$} & \colhead{$\sigma_3$} 
}
\startdata
Emissivity $\epsilon_{W1W2}$ & 0.299 & 0.977 & 0.004 & 0.195 & 0.945 & 0.015 & 0.506 & 0.862 & 0.030 \\
Albedo p$_{W1W2}$            & 0.296  & 0.060 & 0.010 & 0.203 & 0.142 & 0.039 & 0.501 & 0.358 & 0.077  \\
\enddata
\tablenotetext{a}{Best-fit 3-component Gaussian mixture shown in two bottom panels in Figure~\ref{fig:define2}.}
\end{deluxetable}

\subsection{Understanding \atm\ Best Fits \label{sec:FitAnalysis2}} 

Despite good agreement between \atm\ and NEOWISE fitting results, there are statistical 
problems with best-fit models when $\epsilon_{W3} = \epsilon_{W4}$ = 0.90: the per-band 
magnitude residual distributions for all objects are offset from zero by 0.1--0.2 mag 
in most bands. These offsets indicate that the thermal models used are not fully capable of explaining 
WISE data. The distributions of observed objects in WISE color-color diagrams offer 
an efficient way to study possible causes of model deficiencies. We note that 
the use of a full sample to constrain priors for individual objects is known
as Hierarchical Bayesian modeling (e.g., see Chapter 5 in \citealt{zeljkoBook}).

\begin{figure}[th]
\centering
\includegraphics[width=0.9\textwidth, keepaspectratio]{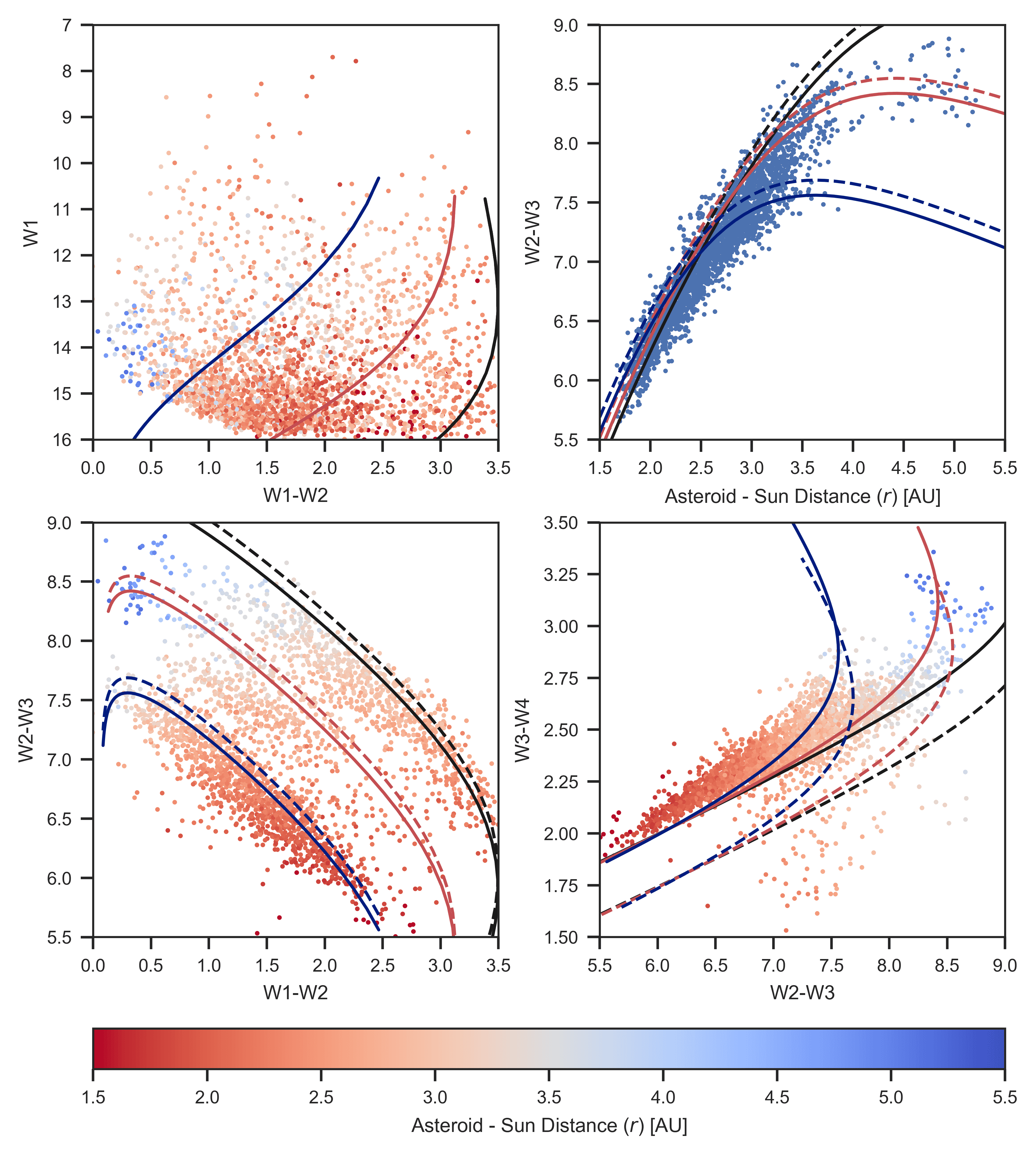}
\caption{The symbols correspond to 2,656 asteroids with $\chi^2 < 3$ (\atm) and 
at least 28 WISE observations. Except for the top right panel, symbols 
are color-coded according to the median asteroid-Sun distance, $r$. 
The top right panel shows the variation of W2-W3 color with $r$. 
The three solid lines in each panel are model tracks produced with three 
fixed pairs of characteristic temperature, $T_1$, and emissivity 
across the W1 and W2 wavelength range, $\epsilon_{W1W2}$ (0.98, 406 K;
0.95, 399 K; 0.86, 391 K; for black, red and blue tracks, respectively). 
The position along each track is controlled by the asteroid-Sun distance 
(models are computed for the range 1.5--5.5 AU). Colors do not depend 
on the object size and distance from the observer; however, the position 
in the W1 vs. W1-W2 color-magnitude diagram (top left) depends on both 
of these quantities and model tracks assume a fiducial asteroid diameter 
of 10 km at a distance of 1 A.U. Emissivity across the W3 and W4 wavelength 
range, parameterized as $\epsilon_{W3}$ and $\epsilon_{W4}$, controls the 
placement of model tracks in the W3-W4 vs. W2-W3 color-color diagram. 
The model tracks shown by solid lines are computed using $\epsilon_{W3} = 0.80$ 
and $\epsilon_{W4} = 0.98$. Model tracks shown by dashed lines correspond 
to the usually adopted values $\epsilon_{W3} = \epsilon_{W4} = 0.90$, and are 
strongly ruled out by the data distribution in the W3-W4 vs. W2-W3 color-color 
diagram. This figure was generated using \href{https://nbviewer.jupyter.org/github/moeyensj/atm/blob/master/notebooks/analysis/analysis.ipynb}{notebooks/analysis/analysis.ipynb}.
\label{fig:CCDs1_1}}
\end{figure}

For each object, we compute the median observed magnitudes in each WISE band and the median observed colors. The resulting color-color diagrams are shown in 
Figure~\ref{fig:CCDs1_1}. 
The position along the model sequences in the W3-W4 vs. W2-W3 color-color 
diagram (see the bottom right panel in Figure~\ref{fig:CCDs1_1}) is by and large 
controlled by the asteroid-Sun distance $r$, while both $r$ and emissivity $\epsilon_{W1W2}$ 
(or, equivalently, albedo) determine the position in the W2-W3 vs. W1-W2 color-color 
diagram (bottom left panel). Note the three clearly delineated observed sequences in this 
diagram. Their finite width is influenced by both the distribution of 
$\epsilon_{W1W2}$ values, and the distribution of observational phase 
angles (increase in $\epsilon_{W1W2}$ and/or phase angle moves 
model sequences from the bottom left to the top right). Note also that 
the W2-W3 and W3-W4 colors become redder as $r$ increases, while the 
W1-W2 color becomes redder for decreasing $r$ (because of the increasing 
relative contribution of the blue reflected component).

The placement of model tracks in the W3-W4 vs. W2-W3 color-color diagram
is controlled by emissivity across the W3 and W4 wavelength range, parameterized 
as $\epsilon_{W3}$ and $\epsilon_{W4}$, while the position along the tracks is
controlled by the asteroid's temperature (itself controlled by $T_1$ and the
asteroid-Sun distance). The three model tracks shown by dashed lines
correspond to $\epsilon_{W3} = \epsilon_{W4} =0.90$, and are strongly ruled out 
by the data: they don't overlap the majority of data points.

A small fraction of data points (about 3\%) with very blue W3-W4 colors could be, 
at least in principle, either measurement outliers, or interesting objects with 
unusual $\epsilon_{W3}$ and  $\epsilon_{W4}$ values. However, all of them are 
extremely bright in W3 (W3$<$1.8) and essentially represent the top 3\% 
brightest sources. Hence, it is likely that their peculiar behavior in the W3-W4 
vs. W2-W3 color-color diagram is simply due to overestimated W3 flux by about 
0.5 mag. This conclusion is consistent with the recently reported flux corrections for
saturation in the W3 band by \cite{2018arXiv181101454W}. 

The only way to move model tracks to reach the data distribution is to assume
$\epsilon_{W4} > \epsilon_{W3}$. After some experimentation, we 
adopted $\epsilon_{W4} \approx 1.22 \epsilon_{W3}$. The required scale 
of $\epsilon$ is only set when W1 data is added because the W1 band includes
a significant contribution from the reflected light component (proportional to $(1-\epsilon)D^2$, 
rather than to $\epsilon D^2$ for thermal emission component, as discussed earlier).  

We investigated three pairs of values: $(\epsilon_{W3}, \epsilon_{W4})$ = (0.80, 0.98),
(0.76, 0.93) and (0.70, 0.86). All three pairs produce model tracks in Figure~\ref{fig:CCDs1_1} 
that are in a much better agreement with data than $\epsilon_{W3} = \epsilon_{W4} = 0.9$. 
The model tracks could be moved further to completely overlap the data distribution
by adopting $\epsilon_{W4} \approx 1.34 \epsilon_{W3}$ (e.g., $\epsilon_{W3}=0.73, \epsilon_{W4}=0.98$). 
We leave detailed investigation of such models to future work. 

While in all three cases the agreement between the data and the models in color-color diagrams 
is essentially unchanged, the values of best-fit diameters change in inverse proportion to the values of $\epsilon$.
Relative to the best-fit diameters for $(\epsilon_{W3}, \epsilon_{W4})$ = (0.80, 0.98)
case, which on average agree with the NEOWISE values to better than 1\%, the
sizes for (0.76, 0.93) case are on average 4\% larger, and 10\% larger for (0.70, 0.86) case. 
Note that best-fit diameters scale with $\epsilon$ faster ($\propto \epsilon^{-0.7}$)
than the naively expected $1/\sqrt{\epsilon}$ due to non-linear fitting effects. 
We chose $(\epsilon_{W3}, \epsilon_{W4})$ = (0.80, 0.98) for the rest of analysis here
because of agreement with the NEOWISE values but we emphasize that these
fiducial diameter values could easily have a systematic error exceeding 10\%
simply due to incorrect priors for $\epsilon_{W3}$ and $\epsilon_{W4}$.  
It is plausible that these biases could be larger for the  
$\epsilon_{W4} \approx 1.34 \epsilon_{W3}$ case. 

This simple analysis illustrates how data can constrain model parameters,
and also justifies our updated choice of priors for $\epsilon_{W3}$ and $\epsilon_{W4}$.
We now proceed with the analysis of best-fit model parameter distributions,
their relationship to data properties, and derive an approximate size estimator
that can be applied to the majority of objects in WISE sample.  

\begin{figure}[th]
\centering
\includegraphics[width=0.9\textwidth, keepaspectratio]{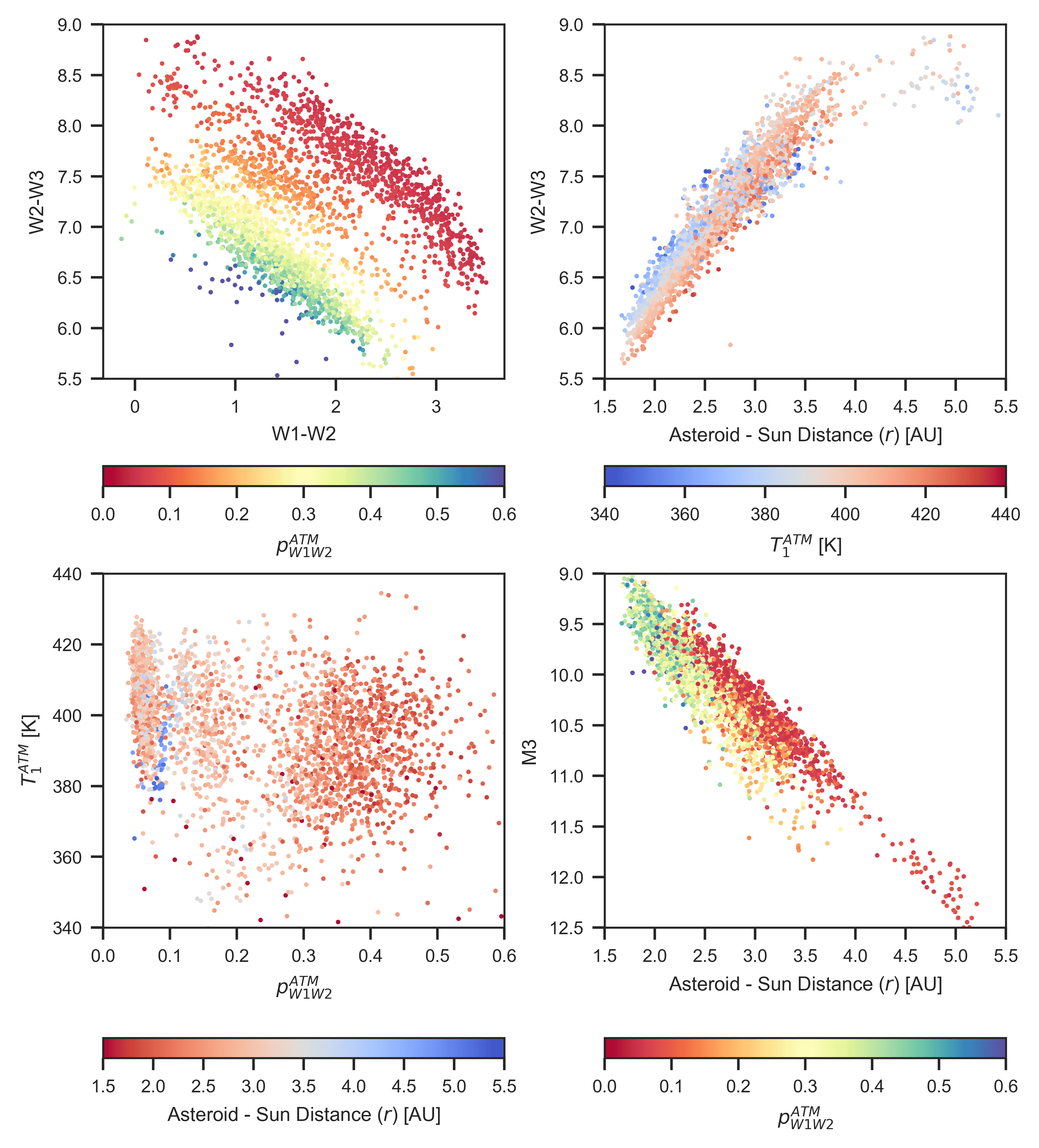}
\caption{The symbols correspond to 2,656 asteroids with $\chi^2 < 3$ (\atm) and 
at least 4 observations in each band. The top left panel shows the W2-W3 vs. W1-W2 
color-color diagram, with symbols color-coded by the best-fit values
of $\epsilon_{W1W2}$. The three observed sequences clearly correspond
to different taxonomic groups. The top right panel shows that the W2-W3
color is by and large controlled by the asteroid-Sun distance, $r$, while
characteristic temperature, $T_1$, has only a secondary influence due 
to its small dynamic range. The bottom left panel shows a weak 
correlation between albedo (or equivalently $\epsilon_{W1W2}$) and 
$T_1$ (the albedo distribution is tri-modal, see the bottom right
panel in Fig.~\ref{fig:define2}). The bottom right panel shows a correlation 
of “absolute” magnitude (corrected for the asteroid size and 
asteroid-observer distance) with $r$. The preponderance of low-albedo 
objects at large $r$ reflects the structure of the asteroid belt. This figure was generated using \href{https://nbviewer.jupyter.org/github/moeyensj/atm/blob/master/notebooks/analysis/analysis_W3.ipynb}{notebooks/analysis/analysis\_W3.ipynb}.
\label{fig:CCDs2}}
\end{figure}

\subsection{An Approximate Single-band W3-based Size Estimator \label{sec:W3approx}} 

The availability of WISE measurements varies greatly with bandpass and only
a small minority of asteroids have data in all four bands. The majority of 
asteroids have data only in W3 and W4 bands, or in the W3 band alone. The best
coverage is in W3 band. The NEOWISE team developed an elaborate set of different 
modeling approaches depending on which bands have data (for a concise summary,
see \S3 in \citealt{NM2018b}). Here we propose a much simpler two-step calibration 
and modeling scheme: 
\begin{enumerate}
\item In the first step, size, temperature parameter and infrared emissivity 
$\epsilon_{W1W2}$ are fit for several thousand asteroids with WISE data in all four bands.
The subset of these objects with direct size measurements is used to calibrate 
and validate NEATM model parameters and priors (not done here but implicitly used since 
the NEOWISE team validated their best-fit sizes using about 100 such objects,
which we reproduce here with a subpercent bias). 
\item In the second step, only W3 measurements are used to estimate the object's
size, and the method is calibrated and validated using the four-band sample from
the first step. 
\end{enumerate}
Hence, for studies requiring large samples of objects with size estimates (e.g.,
tens of thousands), the second step can provide a uniform dataset with well 
understood random and systematic size uncertainties from the 4-band sample,
which is in turn calibrated using objects with direct size measurements. 

The top left panel in Figure~\ref{fig:CCDs2} shows the distribution of objects
in the W2-W3 vs. W1-W2 color-color diagram, with symbols coded by the 
near-IR albedo derived from the best-fit emissivity $\epsilon_{W1W2}$. We use
model with $\epsilon_{W3} = \epsilon_{W4} = 0.9$ because of direct comparison
with NEOWISE results further below. As
evident, the three sequences displayed by data distribution are related to 
albedo, and to its tri-modal distribution shown in Figure~\ref{fig:define2}.
As was demonstrated in Figure~\ref{fig:CCDs1_1}, the position along 
each sequence is by and large controlled by the asteroid temperature, 
which itself is controlled by the best-fit $T_1$ and asteroid-Sun distance. 
These relations are illustrated in the top right panel in Figure~\ref{fig:CCDs2}.
The distribution of best-fit $T_1$ is rather narrow and only weakly correlated
with near-IR albedo, as shown in the bottom left panel in Figure~\ref{fig:CCDs2}.

As expected from eq.~\ref{eq:T1}, $T_1$ does not strongly depend on asteroid-Sun distance. 
Given this independence and a rather narrow distribution of $T_1$, it is possible
to derive an approximate but relatively precise predictor for the asteroid flux,
given the observing geometry (asteroid-Sun distance, $r$, and asteroid-observer
distance, $\Delta$). We first define a ``pseudo-absolute'' magnitude that accounts
for the dependence of observed flux on asteroid size ($\propto D^2$) and its 
distance from the observer ($\propto \Delta^{-2}$),
\eq{
\label{eq:absmag}
        M = W + 5\,\log \left ( \frac{D}{\rm km} \right ) - 5\,\log \left (\frac{\Delta}{\rm A.U.} \right ). 
}
The variation of this ``pseudo-absolute'' magnitude constructed using WISE 
W3 band data with asteroid-Sun distance is shown in the bottom right panel in 
Figure~\ref{fig:CCDs2}. At a given asteroid-Sun distance, the distribution of
$M3$ is rather narrow because both $T_1$ and $\epsilon_{W3}$ distributions 
are narrow. 

\begin{figure}[th]
\centering
\includegraphics[width=0.9\textwidth, keepaspectratio]{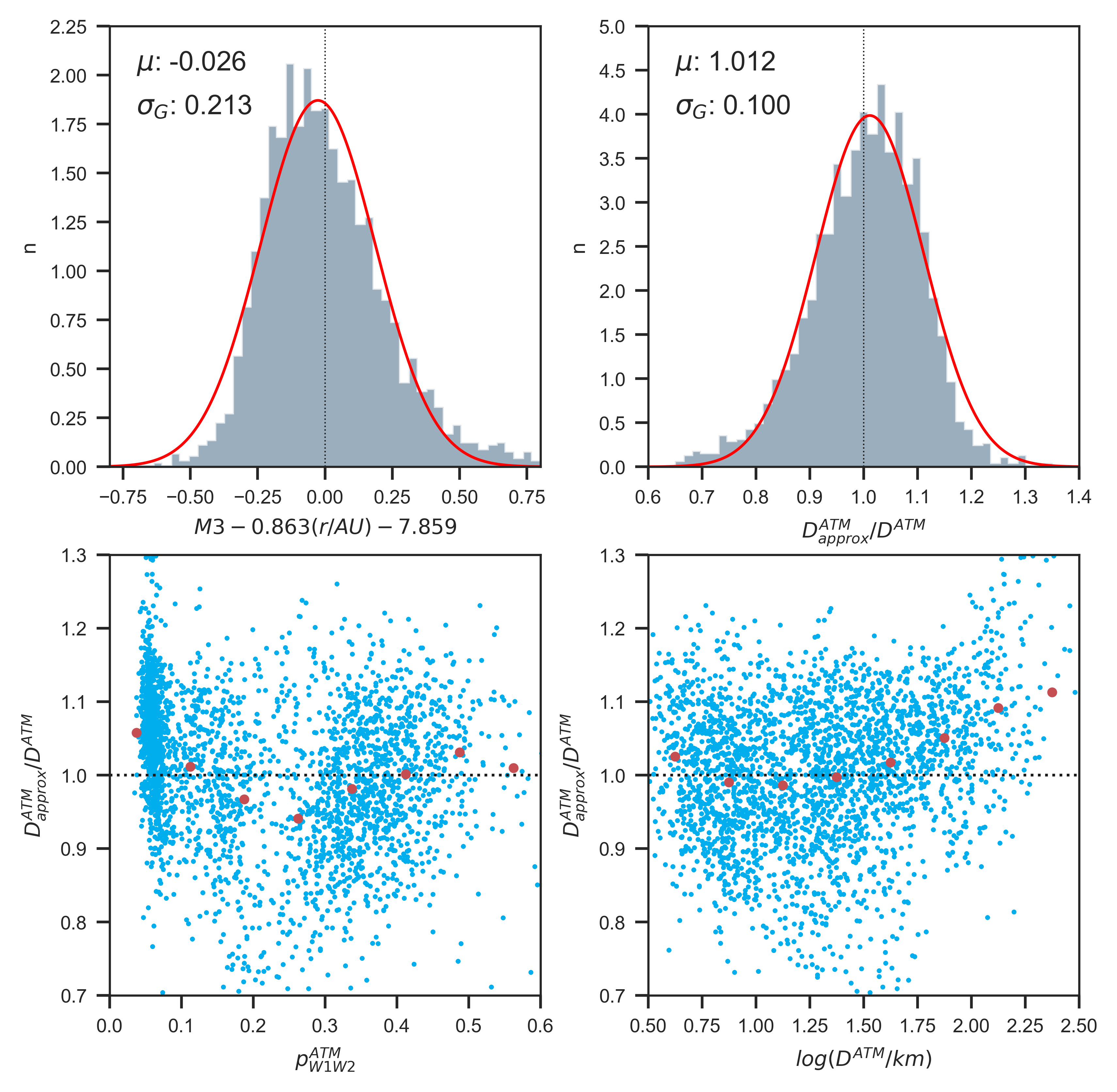}
\caption{The top left panel shows a histogram of residuals 
after a straight line is fit to the 
“absolute” magnitude vs. the asteroid-Sun distance relationship 
shown in the bottom right panel in Fig.~\ref{fig:CCDs2}. The histogram in
the top right panel shows the distribution of the ratio of an 
approximate asteroid diameter estimate based on W3 measurements
alone and the best-fit \atm\ values based on all four WISE bands.
The red lines are single Gaussian fits, with their mean and standard 
deviation shown in each panel. Note that single-band estimates
match 4-band estimates with a scatter of 10\% and a bias of 
1\% (this scatter is not dominated by the formal random uncertainties of 
single-band diameter estimates). The bottom two panels illustrate systematic 
uncertainty in this approximate estimate as a function of albedo 
and diameter. The large symbols are the median values for bins along 
the horizontal axis. This figure was generated using \href{https://nbviewer.jupyter.org/github/moeyensj/atm/blob/master/notebooks/analysis/analysis_W3.ipynb}{notebooks/analysis/analysis\_W3.ipynb}.
\label{fig:appDH3}}
\end{figure}

A linear fit to $M3$ as a function of asteroid-Sun distance,
\eq{
            M3 = 0.863\, \left( \frac{r}{\rm A.U.} \right )+ 7.859. 
}
is adequate and matches observed values with a scatter of 0.21 mag,
as shown in the top left panel in Figure~\ref{fig:appDH3}. Using this fit 
and the definition of $M3$ given by eq.~\ref{eq:absmag}, we derive an
approximate estimator for asteroid size
\eq{
\label{eq:Dapp}
 5\, \log \left( \frac{D_{approx}}{\rm km} \right ) = 7.589 + 0.863\, \left( \frac{r}{\rm A.U.} \right ) - W3 +  5\,\log \left ( \frac{\Delta}{\rm A.U.} \right ). 
}

As shown in the top right panel in Figure~\ref{fig:appDH3}, this single-band
estimator matches best-fit diameters based on data in all four bands with 
a scatter of only about 10\%. This scatter is primarily due to scatter in unknown
$T_1$ and $\epsilon_{W3}$ around their typical values implicitly assumed in eq.~\ref{eq:Dapp}. 
For the same reason, systematic errors are correlated with albedo, at about the 5\%
level as shown in the bottom left panel in Figure~\ref{fig:appDH3}. We also
find a bias of about 10\% for the largest objects (bottom right panel).

The usefulness of the approximate size estimator given by eq.~\ref{eq:Dapp} is primarily
that it can be applied to a much larger sample of objects than the 4-band
sample discussed above. We computed the median W3 magnitude for  
128,660 unique objects, that also have NEOWISE size estimates, using about 
2.3 million W3 measurements. The formal size uncertainties based on 
scatter in observed W3 magnitudes and an intrinsic variability of 15\% are 
in the range 2-6\% for objects with W3 $<8$ and about 10\% at the
sample faint limit (W3$\sim$10).  

\begin{figure}[th]
\centering
\includegraphics[width=0.9\textwidth, keepaspectratio]{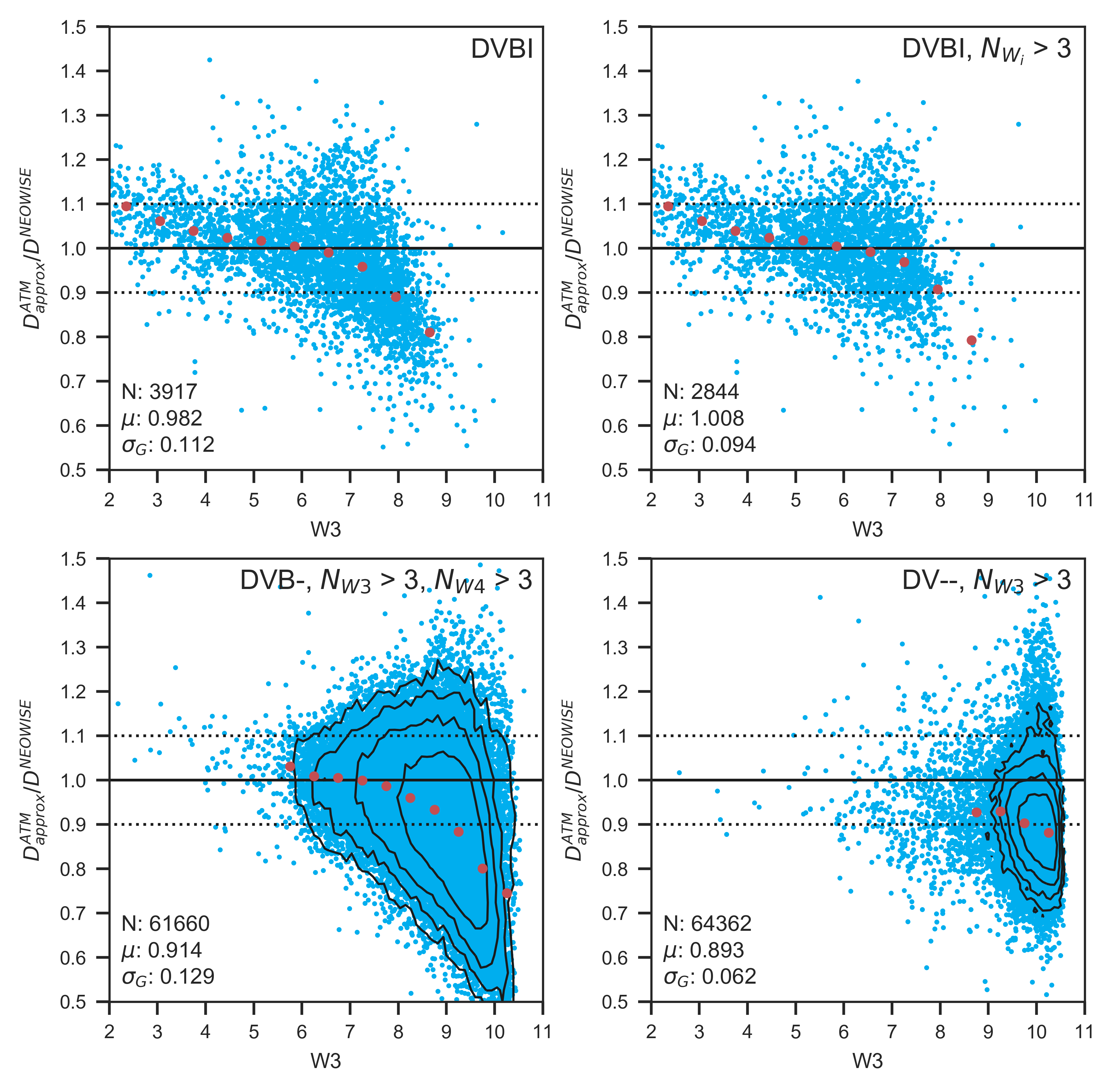}
\caption{
The panels show the ratio of W3-based size estimate and the 2016 NEOWISE
estimate for subsamples of objects selected by their NEOWISE fitting code and
the number of observations (symbols in top two panels and contours in bottom
two panels due to much larger subsamples), as a function of the median observed
W3 magnitude. The large symbols show the median values of size ratio in bins 
of W3. The three numbers in each panel list the subsample size, the median
value of size ratio, and its robust (interquartile based) standard deviation. 
Top panels show ``DVBI'' sample, without (left) and with (right) a limit on the
minimum number of observations in all four WISE bands. Bottom panels show
``DVB-'' subsample (left) and ``DV--'' subsample (right), with a limit on the
minimum number of observations as shown next to the label. This figure was generated using \href{https://nbviewer.jupyter.org/github/moeyensj/atm/blob/master/notebooks/analysis/analysis_W3.ipynb}{notebooks/analysis/analysis\_W3.ipynb}.
\label{fig:Dbiases}}
\end{figure}

The two size estimators are compared in Figure~\ref{fig:Dbiases}, as a function 
of the NEOWISE fit code. They agree on average with a scatter of about 10\%, and
without appreciable biases for the high data quality ``DVBI'' subsample (as 
expected, as this subsample is closely related to the training sample). However, the 
two by far largest subsamples, ``DVB-'' and ``DV--'' show biases exceeding 10\%, 
and these biases have different behavior for different fit codes. Given that we apply
a single estimator, it is likely, although not certain, that biases are introduced by 
the NEOWISE size estimator. As a possible clue, we find that the bias increases with the 
formal uncertainty in median W3 magnitude. 
 
We note that because of a strong correlation between the median W3 flux and 
asteroid size (because of the finite dynamic range of observed distances), the biases with
respect to the median W3 flux propagate to biases with respect to the
object size and thus may introduce biases when estimating size distributions.

\section{Comparison of WISE-based model parameters and SDSS data \label{sec:WISESDSS}}


In this section, we match the ``gold'' sample of 7,359 best-observed
asteroids from the WISE dataset (see \S\ref{sec:WISEdtsel} for selection
criteria) to asteroids with optical observations listed in the 
Sloan Digital Sky Survey Moving Object Catalog (hereafter 
SDSSMOC; \citealt{MOC2001, MOC2002, 2002AJ....124.1776J, MOC2008}). Following \cite{2012ApJ...745....7M}, 
we aim to study correlations between optical and infrared properties, 
such as colors and albedo. In addition, we quantitatively explore the
color vs. albedo correlation and develop an approximate method to
estimate asteroid size from optical data alone. 

The 4th SDSSMOC release\footnote{The 4th Release of SDSSMOC is
available from \url{http://faculty.washington.edu/ivezic/sdssmoc/sdssmoc.html}} 
lists astrometric and photometric data for 471,569 moving objects
observed by SDSS prior to March 2007. Of those, 220,101 objects are
linked to 104,449 unique objects with orbits. A match based on object designation 
to 7,359 objects with WISE-based fits yields 1,574 objects.

In the remainder of the analysis here, we use optical absolute magnitude,
$H$, based on SDSS measurements (field 47 in SDSSMOC) because it represents
an observationally uniform dataset, and because the values obtained from Minor Planet Center 
were found to have errors (both biases and random scatter) at the level of a few tenths of magnitude
(for detailed discussion, see \S2.3 in \citealt{MOC2008}).  However, we note that the net offset 
between the SDSS and MPC values of $H$, reported by \cite{MOC2008},  disappears when using 
the June 2018 version of the MPC catalog MPCORB.DAT, and the scatter is reduced to 0.22 mag.

\subsection{Estimates of optical albedo from WISE-based best-fit sizes}

Using SDSS-based absolute magnitude $H$ and WISE-based best-fit size
$D$, we estimate optical albedo using eq.~\ref{eq:pV}, repeated here for convenience
\eq{
\label{eq:pV2}
      p_V = \left({1329 \, {\rm km}  \over D}\right)^2 \, 10^{-0.4 H}. 
}
This estimate implies that $p_V$ is a free fitting parameter whose 
prior is decoupled from emissivity and albedo values at IR wavelengths
probed by WISE. If some prior data implied a strong emissivity/albedo
relationship across the entire probed wavelength range (e.g., when assuming
a constant unknown value of emissivity $\epsilon$), then a joint fit would be 
more appropriate and $D$ would be constrained by both optical and infrared data. 
Given that our knowledge of the emissivity vs. wavelength curve for 
individual objects is usually poor, and that observed emissivity values 
span a much smaller dynamic range than albedo values, it is better
to first estimate size at wavelengths where thermal emission dominates total observed 
flux, and then use that best-fit size to estimate albedo at wavelengths were
reflected light dominates total observed flux.

\subsection{Optical albedo-color correlation}

\begin{figure}[th]
\centering
\includegraphics[width=0.9\textwidth, keepaspectratio]{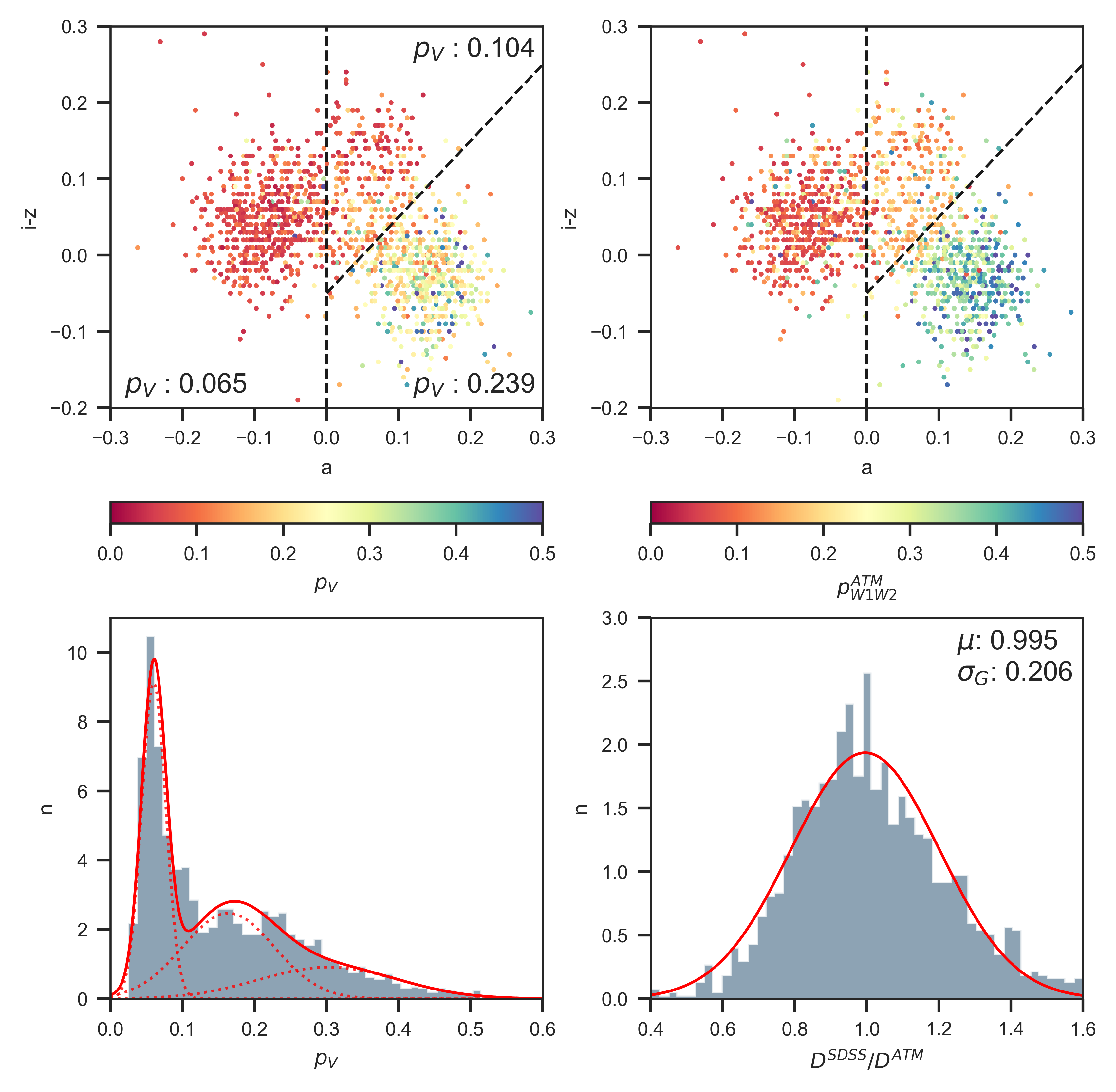}
\caption{The top panels show the $a$ vs. $i-z$ SDSS color-color diagram,
where a color is defined as $a=0.89*gr + 0.45*ri - 0.57$ \citep{MOC2001}, for 1,644 
asteroids from SDSS MOC4 catalog that also have WISE-based distances and
IR albedo estimated with \atm. The symbols are color-coded by WISE-based
IR albedo (right) and by V-band albedo (left) obtained using WISE-based
diameter and SDSS-based absolute magnitude. The vertical dashed line
shows the separation between C and S taxonomic classes from \cite{MOC2001}, while 
the solid line is a separator between low-albedo and high-albedo objects
derived here. The histogram in the bottom left panel shows the distribution 
of V-band albedo and the solid lines show the best-fit 3-component 
Gaussian mixture (fit to individual data points). The darkest component
is centered on $p_V=0.06$. The histogram in the bottom right panel shows
the distribution of the ratio of an approximate asteroid diameter estimate 
based on SDSS measurements alone and the best-fit \atm\ values based on 
all four WISE bands. The SDSS estimate is based on three color-assigned 
values of $p_V$, listed in the top left panel. The mean and 
standard deviation listed in the bottom right panel show that SDSS-based
diameters match WISE-based best-fit \atm\ values with a bias of 
0.5\% and a scatter of 21\%. This figure was generated using \href{https://nbviewer.jupyter.org/github/moeyensj/atm/blob/master/notebooks/analysis/analysis_SDSS.ipynb}{notebooks/analysis/analysis\_SDSS.ipynb}.
\label{fig:SDSS}}
\end{figure}

Following Figure 5 from \cite{2012ApJ...745....7M}, the top two panels 
in Figure \ref{fig:SDSS} show the SDSS $a$ color vs. $i-z$ color-color 
diagram, with symbols color coded using the V band albedo from
eq.~\ref{eq:pV2} (left) and IR albedo derived from best-fit
$\epsilon_{W1W2}$ (right). The $a$ color is the first principal axis 
for the asteroid distribution in the SDSS $r-i$ vs. $g-r$ color-color
diagram, defined as \citep{MOC2001}
\eq{
     a = 0.89 (g-r) + 0.45 (r-i) - 0.57, 
} 
and enables easy separation of C type asteroids ($a<0$) from S type
asteroids ($a>0$). 

As discussed by \cite{2012ApJ...745....7M}, the addition of IR albedo 
improves the definition of taxonomic regions, and presumably of subsets
of asteroids with different optical albedo distributions, in the
optical $r-i$ vs. $g-r$ color-color diagram. Motivated by the
morphology of diagrams shown in the top two panels, in addition
to the $a=0$ separator, we add the line $i-z = a -0.05$ and separate
the sample intro three subsamples. The median optical albedo for 
these subsamples is listed in the top left panel in Figure
\ref{fig:SDSS}. The medians and robust standard deviations are 
listed in Table~\ref{tab:pV}.

\begin{deluxetable}{l|l|l}[t]
\tablecaption{The median and robust standard deviation for the visual albedo, $p_V$. \label{tab:pV}}
\tablehead{
\colhead{Color-selected sample} & \colhead{Median} & \colhead{St.dev.$^a$}
}
\startdata
$a<0$                          &  0.065  &  0.029  \\
$a>0$ \& $i-z > a - 0.05$      &  0.104  &  0.065  \\ 
$a>0$ \& $i-z < a - 0.05$      &  0.239  &  0.089  \\ 
\enddata
\tablenotetext{a}{The robust standard deviation is estimated using interquartile range.} 
\end{deluxetable}

\subsection{An Approximate SDSS-based Size Estimator \label{sec:SDSSapprox}} 

Asteroid size can be estimated from optical data alone by transforming
eq.~\ref{eq:pV2} into
\eq{
\label{eq:pVD}
      D = 1329\, {\rm km} \, {10^{-0.2 H} \over \sqrt{p_V}}
}
and adopting a median albedo. Due to the large dynamic range of visual
albedo, the scatter of such size estimates around the true values is 
large, about 50-60\%, and non-Gaussian. However, the separation
of asteroids using optical colors into three subclasses, each with a
much narrower albedo distribution than for the whole sample, greatly improves 
such estimates. We use the following simple algorithm to assign $p_V$:
\begin{align}
\label{eq:pVcolors}
    p_V = 0.065   &  \,\,\,\, {\rm if} \,\,\,\,  a<0                                          \\  \nonumber 
    p_V = 0.104   &  \,\,\,\, {\rm if} \,\,\,\,  a>0 \,\,\,\, {\rm and} \,\,\,\,  i-z > a - 0.05   \\   \nonumber 
    p_V = 0.239   &  \,\,\,\, {\rm if} \,\,\,\,  a>0 \,\,\,\, {\rm and} \,\,\,\,  i-z < a - 0.05 
\end{align}
As shown in the bottom right panel in Figure~\ref{fig:SDSS}, 
such SDSS-based size estimates match WISE-based estimates with a
scatter of only 21\%, and a nearly Gaussian error distribution. 
For reference, single-band W3-based estimates have an intrinsic 
precision of about 10\%, so the size estimates based on WISE W3 data
are about twice as precise as the SDSS-based estimates (and have similar 
systematic errors because the latter are calibrated using the former). 

We note that the above scatter of 21\% must be at least partially due to the 
variability-induced scatter in single-epoch SDSS estimates of $H$. When
SDSS estimates are replaced by the June 2018 MPC values of $H$, the 
size scatter relative to WISE-based estimates reduces to 17\%. This 
reduction is consistent with a variability-induced scatter of 12\%, 
which is not too dissimilar from the inferred $\Sigma$=0.15 mag
deduced from $\chi^2$ analysis of modeling residuals for WISE fluxes.

\subsection{Selection of M type asteroids using WISE-based best-fit parameters} 

The joint analysis of optical and infrared properties discussed above is focused
on objects with ``typical'' properties. It shows a good correlation between WISE-based
best-fit infrared albedo and optical colors measured by SDSS. Such correlations provide
support that infrared emission models and best-fitting parameters are robust because
the two datasets are essentially independent. Given this independence,  we can also improve 
our understanding of outliers in each dataset. As discussed below, we select a judicious
subsample of outliers using only IR parameters and show that their optical color distribution 
is different than that for the whole sample. This fact further demonstrates that infrared best-fit
model parameters are robust -- if instead IR outliers were random measurement or modeling 
failures, their optical color distribution would not differ from that for the whole sample. 

\cite{2014ApJ...785L...4H} argued that the best-fit IR albedo ($p_{IR}$ in the WISE context, $p_{W1W2}$ in the \atm\ case) and beaming 
parameter ($\eta$) can be used to select metallic asteroids (M taxonomic type). Their 
main argument is that objects with high radar albedo values, indicative of metallic
objects, display a very narrow distribution of IR albedo ($p_{W1W2} \sim 0.2$), while a larger 
fraction of objects with unusually high beaming parameter values are seen in the same 
albedo range. Therefore, objects with large $\eta$ and $p_{W1W2} \sim 0.2$ are good
candidates for metallic asteroids. Since WISE data are available for orders of 
magnitude more objects than radar observations, and metallic objects are 
interesting in many ways (for discussion see \citealt{2014ApJ...785L...4H}), 
it is prudent to critically examine this method. 

Due to degeneracy discussed in \S\ref{sec:emittedFlux}, \atm\ fits only for temperature 
parameter $T_1$ and not for $\eta$.  The relationship between $T_1$ and $\eta$ is given 
by eq.~\ref{eq:T1}; high $\eta$ corresponds to low $T_1$. Therefore, an implication of
analysis from \cite{2014ApJ...785L...4H} is that low $T_1$ objects with $p_{W1W2} \sim 0.2$ 
are good candidates for metallic asteroids. We now examine whether IR data discussed
here suggest that such outliers exist, and if so, whether they have distinct optical colors. 

\begin{figure}[th]
\centering
\includegraphics[width=0.9\textwidth, keepaspectratio]{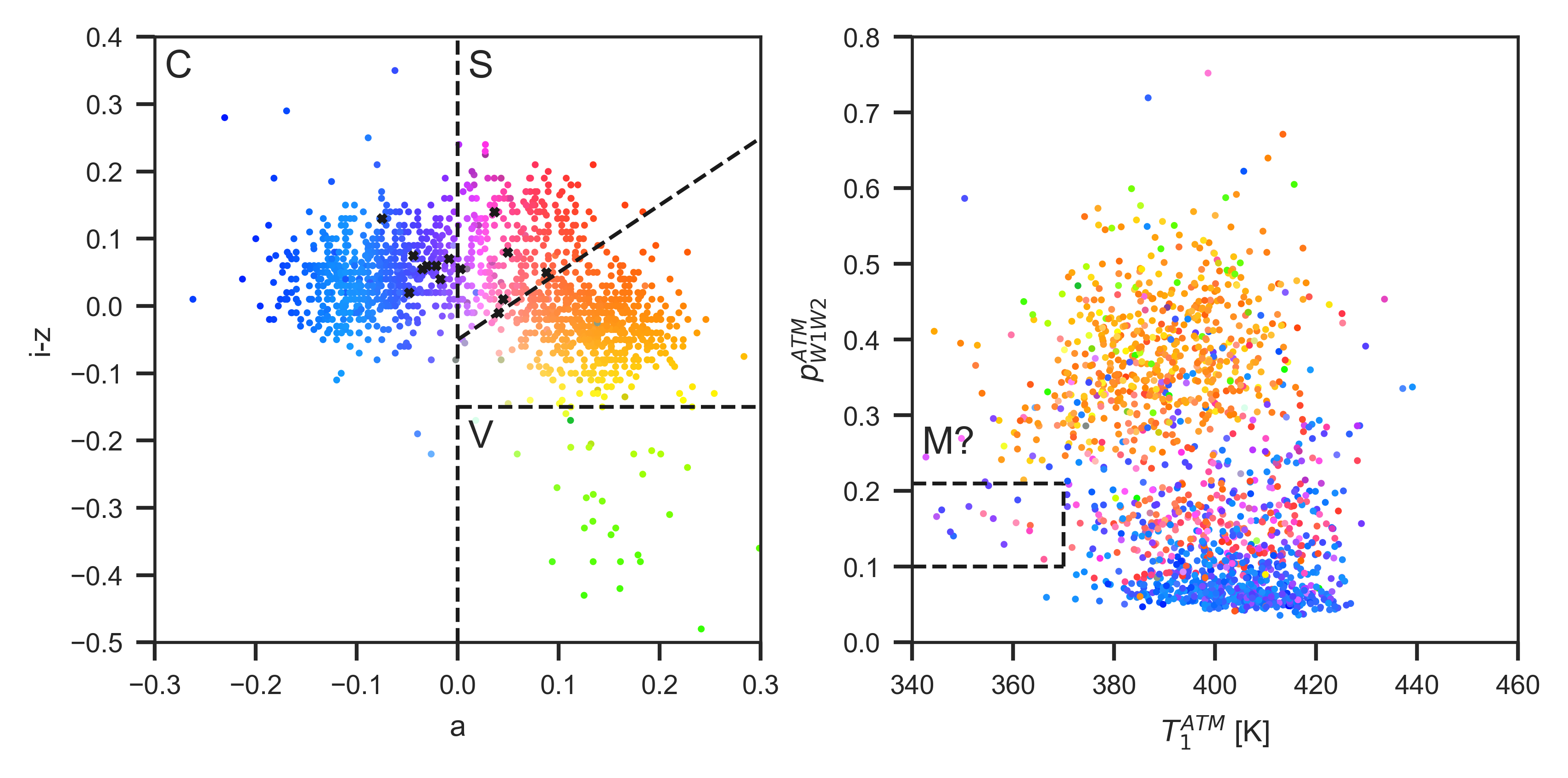}
\caption{The colored symbols in the left panel show the $a$ vs. $i-z$ SDSS 
color-color diagram for the same objects as in Figure~\ref{fig:SDSS}.  
The symbols' color code is two-dimensional,
according to $a$ and $i-z$ colors (for algorithmic 
details see \citealt{MOC2002}). 
The same color coding is used in the right panel to visualize the correlation
of optical colors and WISE-based best-fit values of IR albedo ($p_{W1W2}$) and 
temperature parameter $T_1$. It is easy to discern that, for example, 
objects with $i-z  < -0.15$ have high IR albedo, while objects with $a<0$
have predominantly low IR albedo. The dashed lines in the right panel outline
selection of 14 candidates for metallic asteroids. Their distribution of optical colors
is different from the color distribution for the full sample, as further visualized
by showing them as crosses in the left panel. The dashed lines in the left
panel outline the distribution of main taxonomic classes, as marked in the
panel (C, S and V). This figure was generated using \href{https://nbviewer.jupyter.org/github/moeyensj/atm/blob/master/notebooks/analysis/analysis_SDSS.ipynb}{notebooks/analysis/analysis\_SDSS.ipynb}.   
\label{fig:SDSSWISE}}
\end{figure}

The right panel in Figure~\ref{fig:SDSSWISE} shows that the SDSS-WISE sample does 
contain objects at the low end of $T_1$ range that have $p_{W1W2} \sim 0.2$. By requiring
$340 < T_1 / K < 370$ and $0.1 < p_{W1W2} < 0.21$, we select 14 objects out of 1,644 objects
in the SDSS-WISE sample (there are 32 selected objects out of 2,479 objects in the high-quality 
WISE sample), or about 1\% of the sample. These candidates for metallic asteroids are 
listed in Table~\ref{tab:Mtype}.

\begin{deluxetable}{rrrrrrrrr}[t]
\tablecaption{Candidates for M type (metallic) asteroids$^a$. \label{tab:Mtype}}
\tablehead{
\colhead{Designation} & \colhead{$g-r$}  & \colhead{$r-i$}  & \colhead{$logD$}  & \colhead{$\sigma_{logD}$}  & 
\colhead{$logT_1$}  & \colhead{$\sigma_{logT}$}  & \colhead{$p_{W1W2}$}  & \colhead{$\sigma_{p}$}  
}
\startdata
  (497) &  0.48 &  0.15 &   4.717 &            0.024 &     2.542 &              0.009 &       0.141 &                0.015 \\
  (844) &  0.50 &  0.19 &   4.715 &            0.029 &     2.541 &              0.011 &       0.146 &                0.019 \\
(1349) &  0.51 &  0.16 &   4.466 &            0.040 &     2.539 &              0.015 &       0.175 &                0.033 \\
(1546) &  0.52 &  0.19 &   4.468 &            0.029 &     2.546 &              0.011 &       0.180 &                0.022 \\
(1670) &  0.62 &  0.14 &   4.365 &            0.016 &     2.549 &              0.006 &       0.170 &                0.012 \\
(1730) &  0.58 &  0.13 &   4.213 &            0.025 &     2.538 &              0.010 &       0.166 &                0.018 \\
(1732) &  0.63 &  0.21 &   4.397 &            0.028 &     2.560 &              0.011 &       0.155 &                0.019 \\
(1860) &  0.52 &  0.20 &   4.270 &            0.012 &     2.554 &              0.004 &       0.130 &                0.009 \\
(1977) &  0.56 &  0.15 &   4.284 &            0.021 &     2.552 &              0.008 &       0.164 &                0.016 \\
(2294) &  0.61 &  0.15 &   4.206 &            0.017 &     2.557 &              0.006 &       0.158 &                0.014 \\
(2407) &  0.52 &  0.17 &   4.417 &            0.021 &     2.550 &              0.008 &       0.207 &                0.019 \\
(2573) &  0.61 &  0.17 &   4.325 &            0.020 &     2.564 &              0.007 &       0.110 &                0.011 \\
(2904) &  0.49 &  0.19 &   4.209 &            0.017 &     2.557 &              0.006 &       0.188 &                0.015 \\
(4813) &  0.56 &  0.24 &   4.256 &            0.027 &     2.560 &              0.011 &       0.147 &                0.018 \\
\enddata
\tablenotetext{a}{$g-r$ and $r-i$ are SDSS colors. Diameter $D$ is in meters, $T_1$ in Kelvin.} 
\end{deluxetable}

As the two-dimensional color scheme\footnote{For Python code, see \url{http://www.astroml.org/book\_figures/chapter1/fig\_moving\_objects\_multicolor.html}} 
in Figure~\ref{fig:SDSSWISE} illustrates, the IR-selected objects have significantly different optical $a$ color distribution than 
the full sample: the $a$ color mean and standard deviation for selected objects are $-0.02$ and 0.027, 
respectively (see the left panel in Figure~\ref{fig:SDSSWISE}). This difference
demonstrates that these objects are not random outliers in the $p_{W1W2}$ vs. $T_1$ 
diagram and provides support to the hypothesis about metallic asteroids advanced
by \cite{2014ApJ...785L...4H}. Yet, the optical colors of these metallic candidates
are not sufficiently unique for an efficient selection using only optical data -- 
for example, a restrictive selection based on $a$ and $i-z$ colors that selects
only 8 out of 14 WISE-selected candidates (a selection completeness of $\sim$50\%) still results
in only 10\% sample selection purity (that is, there are about 10 times as many 
other objects in the SDSS subsample selected by the cut). 
 
Infrared data are required to efficiently select candidates for metallic 
asteroids. As the right panel in Figure~\ref{fig:SDSSWISE}
shows, $T_1$ selection is more restrictive than $p_{W1W2}$ selection. Therefore, 
in the context of selecting metallic candidates with a hypothetical two-band
survey, having W3 and W4 data would be more useful than W2 and W3 data. On the
other hand, for studies requiring $p_{W1W2}$, W1 or W2 band would be a more 
useful addition to W3 band than the addition of W4 band.

%

\section{Discussion and conclusions}

Motivated by a desire to enable transparency and reproducibility of results, and to
foster collaborative software development, we release \atm, a general tool
for interpreting infrared flux measurements of asteroids. With
adequate infrared data, this tool can be used to
estimate asteroid sizes, constrain asteroid emissivities at infrared
wavelengths, and when optical data are available, also estimate visual 
albedos. The package also includes data files and example Jupyter
Notebooks that can help significantly reduce the time to reproduce 
published results. All the analysis presented here, including all the 
figures,  tables, and catalogs, can be easily reproduced with these
Notebooks.  

We emphasize that our analysis presented here, and corresponding catalogs
with best-fit sizes and other parameters, are far from definitive and 
can be improved in various ways. For example, an analysis of emissivity
in W1 and W2 bands using Hierarchical Bayesian modeling, similar to our
analysis of emissivity in W3 and W4 bands (which could also be improved
by optimizing the adopted $\epsilon_{W4}/\epsilon_{W3}$ ratio), would 
likely further decrease systematic uncertainties. As another example, 
modern machine learning methods could be used to improve our simplistic 
algorithm for assigning visual albedo using optical colors (eq.~\ref{eq:pVcolors}). 
We leave these improvements for future studies by us and the community. 

Nevertheless, our results presented here already yield a number of useful 
conclusions. We show that \atm\ can match the best-fit size estimates for
best-observed objects published in 2016 by the NEOWISE team with a
sub-percent bias and a scatter of only 6\%. Plausible reasons for this
scatter include different outlier rejection algorithms, different
treatments of Kirchhoff's law, and \atm\ accounting for intrinsic
variability, although we cannot exclude other causes. 
Whatever the reason, the discrepancies are encouragingly small. 

Our analysis of various sources of random and systematic size
uncertainties show that for the majority of over 100,000 objects with WISE-based 
size estimates random uncertainties (precision) are about 10\% (using W3-based
estimates calibrated using high-quality 4-band subsample, see
\S\ref{sec:W3approx}), and
systematic uncertainties within the adopted 
model framework, such as NEATM, are in the range 10-20\%. We estimate 
that the accuracy of WISE-based asteroid size estimates is in the
range 15-20\% for most objects, except for unknown errors due to
an inadequate modeling framework (such as spherical asteroid approximation).
Of course, there is no implied 
guarantee of Gaussianity and these statements need to be interpreted
with care. This result is consistent with the
statement that accuracy is about 15\% by \cite{2016PDSS..247.....M},
but somewhat larger than the claim of ``errors better than 10\%'' 
in \cite{2011ApJ...741...68M}. 

The treatment of priors for emissivity $\epsilon(\lambda)$ has a
direct and fundamental impact on biases in best-fit size estimates. We note that 
given this role of $\epsilon(\lambda)$, which is presumably shared
by all members of an asteroid family, resulting systematic errors
will be shared by all family members and thus cannot be seen when
analyzing the per-family scatter in WISE-based optical albedo 
\citep{2018AJ....156...62M}. Studies of such
scatter are insensitive to systematic errors due to incorrect
$\epsilon(\lambda)$ and thus cannot be used to constrain the absolute 
uncertainty, or accuracy, of flux-based size estimates. Only direct
size measurements can enable a full understanding of the accuracy of 
size estimates based on thermal models. 

Our \atm\ results faithfully recover the tri-modal distribution of
$\epsilon_{W1W2}$ emissivity related to taxonomic classes discovered by \cite{2014ApJ...791..121M}. Correlations of SDSS colors and
WISE-based best-fit model parameters indicate the robustness of the
latter, and also give support to the claim that candidate metallic 
asteroids can be selected using best-fit temperature parameter and 
IR albedo \citep{2014ApJ...785L...4H}. However, it should be noted that the condition on priors
$\epsilon_{W1} = \epsilon_{W2} = \epsilon_{W1W2}$, introduced because WISE 
data do not strongly constrain $\epsilon_{W2}$, may not be optimal. 
For example, we did not investigate ansatz $\epsilon_{W2} = k\,
\epsilon_{W1}$, with $k$ different from unity, as we did for 
$\epsilon_{W3}$ and $\epsilon_{W4}$. We noticed some evidence for 
$k < 1$ when both $\epsilon_{W1}$ and $\epsilon_{W2}$ are free fitting
parameters (see \S\ref{sec:use} and Table~\ref{tab:SEDs}). Investigation of the optimal value of $k$ using the full 
high-quality sample and Hierarchical Bayes methodology (as we did
for the $\epsilon_{W4}/\epsilon_{W3}$ ratio, see \S\ref{sec:FitAnalysis2})
may shed new light on the behavior of infrared emissivity/albedo 
(especially if attempted for taxonomic subsamples defined by optical
colors). 

We utilized a correlation between SDSS optical colors and optical
albedo derived using WISE-based size estimates and developed a method
to estimate asteroid sizes with optical data alone, with an
uncertainty of about 21\% relative to WISE-based size estimates. 
When systematic errors are included, this small difference in 
accuracy between IR-based and optical size estimates is further
diminished. This remarkable result bodes well for future optical asteroid surveys,
such as the Large Synoptic Survey Telescope \citep{LSSToverview}, which might deliver
such size estimates for over 5 million asteroids \citep[][and references therein]{2018Icar..303..181J}. 

Nevertheless, we point out that adequate infrared data are crucial
for breaking the degeneracy between emissivity and asteroid size. 
An infrared survey with appropriately placed (at least) three bandpasses,
and sensitivity to match the LSST sample, could
provide a major breakthrough in our knowledge of the emissivity
distribution for asteroid population. We also note that laboratory
measurements of emissivity can greatly contribute to this endeavor
by providing more robust priors for $\epsilon(\lambda)$. Last but
not least, direct asteroid size measurements are of paramount
importance for validating thermal asteroid models and quantitatively estimating 
their intrinsic biases, and they should be greatly encouraged and supported.

\acknowledgments

J. Moeyens and \v{Z}. Ivezi\'{c} acknowledge support from the University of Washington College of Arts and Sciences, Department of Astronomy, and the DIRAC Institute. The DIRAC Institute is supported through generous gifts from the Charles and Lisa Simonyi Fund for Arts and Sciences, and the Washington Research Foundation.

J. Moeyens thanks the LSST Corporation Data Science Fellowship Program, his time as a Fellow has benefited this work.

We thank the organizers of an asteroid modeling workshop at the University
of Washington (October 3-4, 2016), where this project was conceived: LSST
Corporation, NASA NEO Office and the B612 Foundation.

This publication makes use of data products from the Wide-field Infrared Survey Explorer, which is a joint project of the University of California, Los Angeles, and the Jet Propulsion Laboratory/California Institute of Technology, funded by the National Aeronautics and Space Administration.

This work was facilitated though the use of advanced computational, storage, and networking infrastructure provided by 
the Hyak supercomputer system at the University of Washington.

\software{pymc3 \citep{pymc3}, numpy \citep{numpy}, scipy \citep{scipy}, pandas \citep{pandas}, astropy \citep{astropy-1, astropy-2}, matplotlib \citep{matplotlib}, corner \citep{corner}, seaborn \citep{seaborn}, astroML \citep{2012cidu.conf...47V}}

\bibliographystyle{aasjournal}
\bibliography{ref}{}

\appendix

\section{A correction to the quadrature formula for the W3 band} 

\cite{2013AAS...22143905W} has derived simple quadrature formulae that can be used to compute 
in-band fluxes for the four WISE bands from model flux $F_\nu^{ast}(\lambda)$. 
It appears that the provided coefficients for the W3 band can be improved 
(N. Myhrvold, in prep.). The corrected coefficients are available in 
\atm\ package, in method bandpassLambda that can be found in file \href{https://github.com/moeyensj/atm/blob/master/atm/obs/wise.py#L67}{atm/obs/wise.py}. 
We validated new coefficients by comparing the approximate integral
obtained using the quadrature formula to exactly integrated flux for
a $T=100$ K black body. The new coefficients match the exact integral
to better than 0.2\%, while the original coefficients result in a 35\% 
smaller flux. We have verified that the original coefficients for other 
three bands match exact integrals to sub-percent accuracy.

\end{document}